\documentclass[%
superscriptaddress,
preprint,
 amsmath,amssymb,
 aps,
floatfix,
]{revtex4-2}

\usepackage{graphicx}
\usepackage{bm}
\usepackage{siunitx}
\usepackage{xcolor}
\usepackage{mathtools}

\makeatletter
\def\@fnsymbol#1{\ensuremath{\ifcase#1\or \dagger\or *\or \ddagger\or
   \mathsection\or \mathparagraph\or \|\or **\or \dagger\dagger
   \or \ddagger\ddagger \else\@ctrerr\fi}}
\makeatother

\begin{document}

\title{Vibrationally resolved optical excitations of the nitrogen-vacancy center in diamond}

\author{Yu Jin}
\affiliation{Department of Chemistry, University of Chicago, Chicago, Illinois 60637, United States}
\author{Marco Govoni}
\email{mgovoni@anl.gov}
\affiliation{Pritzker School of Molecular Engineering, University of Chicago, Chicago, Illinois 60637, United States}
\affiliation{Materials Science Division and Center for Molecular Engineering, Argonne National Laboratory, Lemont, Illinois 60439, United States}
\author{Giulia Galli}
\email{gagalli@uchicago.edu}
\affiliation{Department of Chemistry, University of Chicago, Chicago, Illinois 60637, United States}
\affiliation{Pritzker School of Molecular Engineering, University of Chicago, Chicago, Illinois 60637, United States}
\affiliation{Materials Science Division and Center for Molecular Engineering, Argonne National Laboratory, Lemont, Illinois 60439, United States}

\date{\today}

\begin{abstract}
A comprehensive description of the optical cycle of spin defects in solids requires the understanding of the electronic and atomistic structure of states with different spin multiplicity, including singlet states which are particularly challenging from a theoretical standpoint. We present a general framework, based on spin-flip time-dependent density function theory, to determine the excited state potential energy surfaces of the many-body singlet states of spin defects; we then predict the vibrationally resolved absorption spectrum between singlet shelving states of a prototypical defect, the nitrogen-vacancy center in diamond. Our results, which are in excellent agreement with experiments, provide an interpretation of the measured spectra and reveal the key role of specific phonons in determining absorption processes, and the notable influence of non-adiabatic interactions. The insights gained from our calculations may be useful in defining strategies to improve infrared-absorption-based magnetometry and optical pumping schemes. The theoretical framework developed here is general and applicable to a variety of other spin defects and materials.
\end{abstract}

\maketitle

\section{\label{sec:intro} Introduction}
Spin defects in semiconductors and insulators have attracted considerable attention in the last decade, as promising platforms to realize quantum technologies~\cite{wolfowicz2021quantumguideline}. For example, it has been shown  that simple point defects such as the negatively charged nitrogen-vacancy (NV$^-$) in diamond~\cite{walker1979optical} may be used as quantum bits (qubits), where the qubit initialization and readout is realized through an optical spin-polarization cycle between the triplet ground state, a triplet excited state  and two shelving singlet states~\cite{robledo2011spin, choi2012mechanism, goldman2015state, thiering2018nvisc}. The ability to initialize and readout the NV$^-$ center in diamond has led to numerous proposals for quantum technology applications~\cite{doherty2013nitrogen, gali2019nvreview}, including quantum sensing~\cite{schirhagl2014nvsensor, barry2020nvmag} and  communication~\cite{childress2013diamond}, and possibly quantum computation~\cite{weber2010quantum, waldherr2014quantum}.

While the optical and magnetic properties of the triplet ground and the first triplet excited state of the NV$^-$ center have been extensively investigated using density functional theory (DFT)~\cite{gali2009nv, gali2011tddft, gali2017nvdjt, alkauskas2014luminescence, ivady2018first, gali2019nvreview, razinkovas2021nvpl, jin2021pl}, robust first-principles predictions of the properties of the singlet shelving states are not yet available. The reason is two-fold: the description of the electronic structure of these singlet states require a higher level of theory than DFT to account for their strongly correlated (multiconfigurational) nature; in addition, the determination of their atomistic structure requires techniques capable of optimizing complex excited state potential energy surfaces (PESs), beyond DFT with constrained occupations ($\Delta$SCF). Important progress has been reported in using high level theories to investigate the electronic structure of the shelving singlets of the NV$^-$ center, at fixed geometries; these theories include many-body perturbation theory ($GW$ and the solution of the Bethe-Salpeter
Equation (BSE)~\cite{ma2010nvgwbse}), quantum chemistry methods, e.g., complete active space self-consistent field (CASSCF)~\cite{bhandari2021casscf}, the diagonalization of effective Hamiltonian derived within the constrained random phase approximation (CRPA)~\cite{bockstedte2018ab}, and a quantum  embedding theory (QDET)~\cite{he2020npj, he2020pccp, he2021jctc, vorwerk2021qdet, huang2022qs, nan2022dct}. However, all these approaches have been limited to the evaluation of vertical excitation energies at given geometries and the PESs of the singlet states, and their vibrationally resolved optical spectra have not been predicted from first principles. In a pioneering work, Thiering and Gali~\cite{thiering2018nvisc} investigated optical transitions and inter-system crossings involving singlet states, based on a model Hamiltonian parameterized by DFT calculations. However, they included  parameters fitted to experiments, e.g., the energy spacing between the singlet states, and overall they obtained a fair agreement between experiments and computed absorption spectra.

With the goal of providing a comprehensive description of the optical cycle of the NV$^-$ center, we investigate the electronic and atomistic structure of the singlet states involved in the optical cycle. We present a general framework based on the implementation of spin-flip time-dependent density function theory (TDDFT)~\cite{wang2004sftddft,wang2005sftddft,li2012sftddft,bernard2012sftddft,casanova2020sf,walker2006efficient,nguyen2019finite} using a plane-wave basis set, which allows for an accurate determination of the excited states PESs. We use both the semi-local functional by Perdew, Burke and Ernzerhof (PBE)~\cite{PBE} and dielectric dependent hybrid (DDH) functionals~\cite{skone2014ddh} and we evaluate analytical forces acting on the nuclei~\cite{hutter2003excited,seth2011sfforces}. By computing many-body electronic states, equilibrium geometries, and phonons of the singlet states, we successfully predict, for the first time, the infrared vibrationally resolved absorption spectrum~\cite{kehayias2013nvexp} between singlet shelving states using the Huang-Rhys (HR) theory~\cite{HR-theory, alkauskas2014luminescence, razinkovas2021nvpl, jin2021pl}. Our results, which are in excellent agreement with experiments, provide an interpretation of the measured spectra and reveal the key role of specific phonons in determining absorption processes, and the notable influence of non-adiabatic interactions. The insights gained from our calculations may be useful in defining strategies to improve infrared-absorption-based magnetometry~\cite{dumeige2013absmag, jensen2014absmag,wickenbrock2016absmag,chatzidrosos2017absmag} and optical pumping schemes. The theoretical framework developed and used here is general and applicable to a variety of other spin defects and materials.

The rest of the paper is organized as follows. We first present our electronic structure calculations of the many-body electronic states of the NV$^-$ center at a fixed geometry, followed by the determination of their PESs. We then discuss electron-phonon coupling and finally present the vibrationally resolved optical absorption spectrum of the spin-defect. We close the paper with a discussion and summary of all the results.

\section{\label{sec:results}Results}

\subsection{\label{subsec:vee}Many-body electronic states and vertical excitation energies}
\begin{figure}[t]
    \includegraphics[width=16cm]{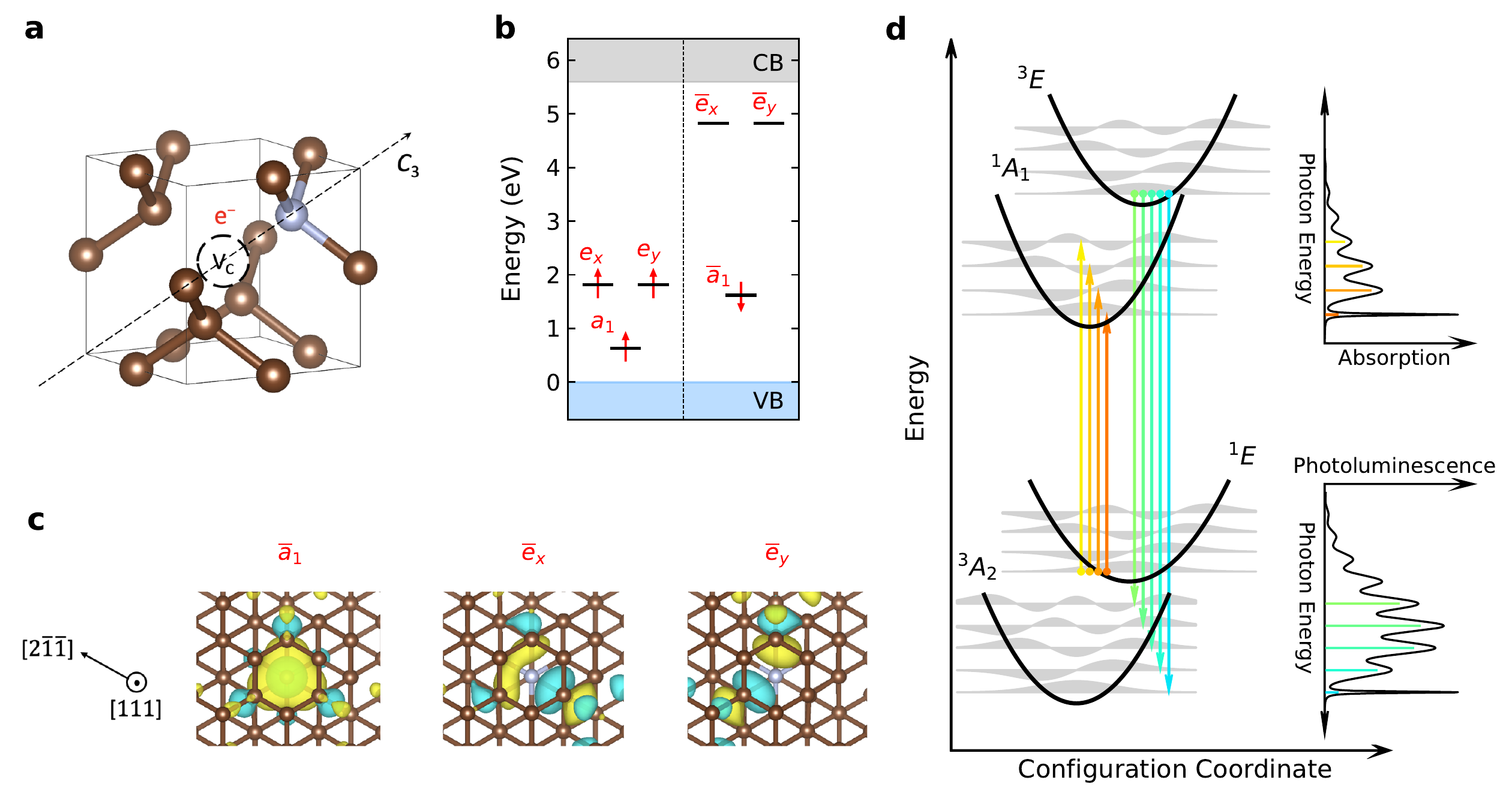}
    \caption{\label{fig:defect-level} \textbf{Description of the NV$^-$ center in diamond.} \textbf{a} Ball and stick representation, with the vacancy depicted as a circle in the middle of the diamond cage, and the carbon and nitrogen atoms represented by brown and gray spheres, respectively. The defect has $C_{3v}$ symmetry, with a threefold rotation axis ($C_3$) parallel to the $\langle111\rangle$ axis of diamond. \textbf{b} Position of the single-particle defect levels in the band gap of diamond, labeled according to the irreducible representation of the $C_{3v}$ group, and computed by spin unrestricted density functional theory calculations with the DDH hybrid functional~\cite{skone2014ddh}. \textbf{c} Isosurfaces of the square moduli of the single particle orbitals associated to the defect levels. The color (yellow/light blue) represents the sign ($+$/$-$) of the orbital. \textbf{d} Schematic diagram illustrating optical processes leading to the photoluminescence (PL) of the $\prescript{3}{}{E} \to \prescript{3}{}{A}_2$ transition and the absorption of the $\prescript{1}{}{E} \to \prescript{1}{}{A}_1$ transition (see text). For ease of graphical representation, the potential energy surfaces are shown as parabolas. Vibrational wavefunctions are schematically shown in gray. Colored arrows represent optical transitions at 0 K. PL and absorption line shapes containing sharp zero-phonon lines and broad phonon side bands are shown as insets.}
\end{figure}

As well known, the NV$^-$ center in diamond is composed of a nitrogen impurity and an adjacent carbon vacancy ($\text{V}_\text{C}$) (see Fig.~\ref{fig:defect-level}). The defect has $C_{3v}$ symmetry, with three orbitals within the band gap of diamond (one $a_1$ and twofold-degenerate $e$ orbitals), localized on three carbon sites in the vicinity of $\text{V}_\text{C}$. Hereafter, we denote the spin up (down) defect orbitals as $a_1$, $e_x$, $e_y$ ($\overline{a}_1$, $\overline{e}_x$, $\overline{e}_y$). The low-lying many-body triplet states are denoted as $\prescript{3}{}{A}_2$ (ground state) and $\prescript{3}{}{E}$  and the singlet states as $\prescript{1}{}{E}$ and $\prescript{1}{}{A}_1$~\cite{doherty2011negatively, maze2011properties}. In the $m_s = 1$ sublevel of the $\prescript{3}{}{A}_2$ ground state, $a_1$, $e_x$, $e_y$ and $\overline{a}_1$ are occupied by four electrons, while $\overline{e}_x$, $\overline{e}_y$ are empty, and its electronic configuration is represented by the Slater determinant $|\overline{e}_x\overline{e}_y\rangle$ in the hole notation. Similarly, phonons modes of the NV$^-$ center are also labeled as $a_1$, $a_2$ and $e$ type according to the $C_{3v}$ point group.

\begin{figure}[t]
\includegraphics[width=10cm]{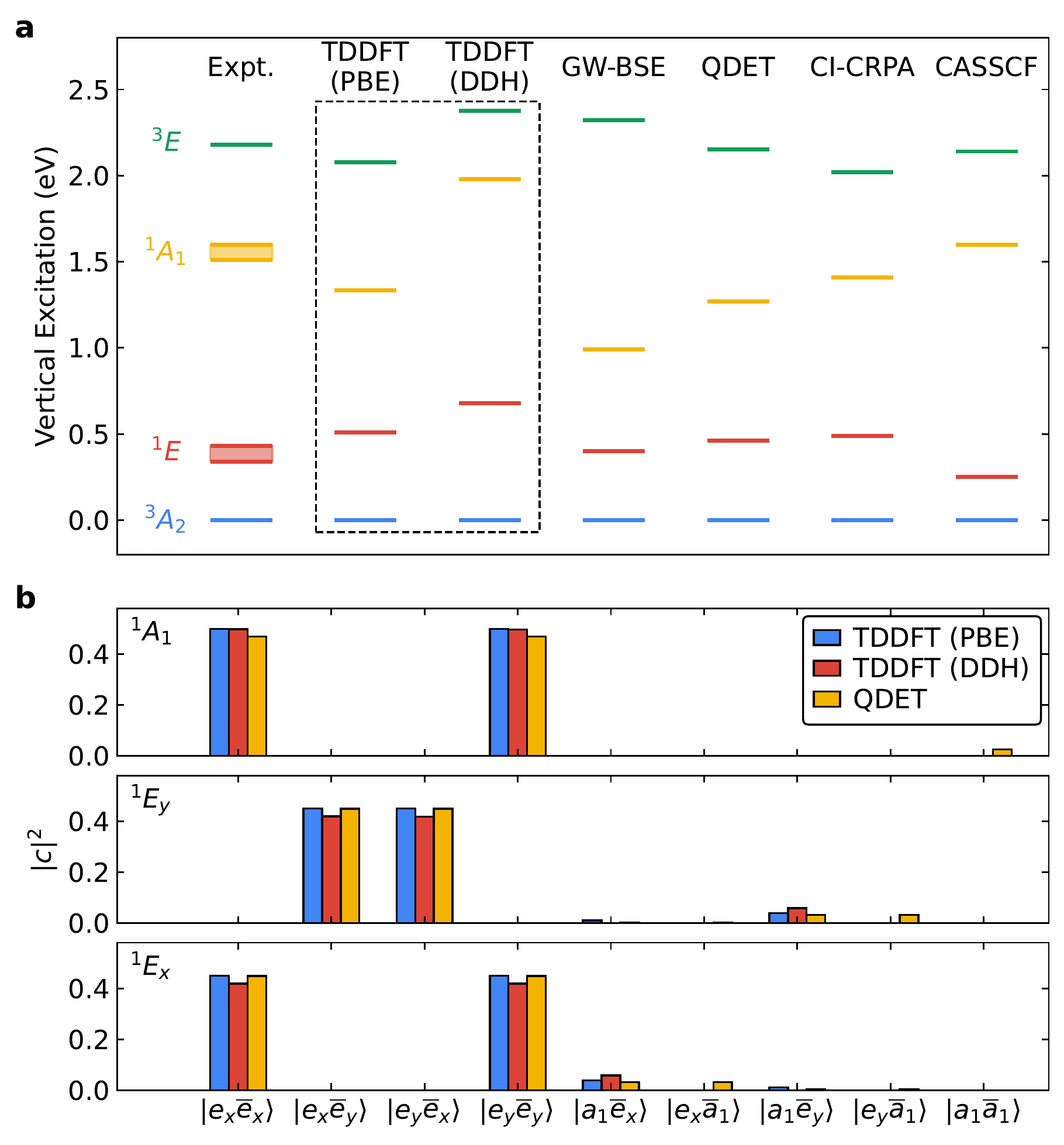}
\caption{\label{fig:vee} \textbf{Many-body electronic states of the NV$^-$ center in diamond.} \textbf{a} Vertical excitation energies (VEEs) of the low-lying many-body electronic states at the ground state geometry, computed using time dependent density functional theory (TDDFT), and using PBE and DDH functionals are shown in the rectangle. Experimentally inferred VEE of the $\prescript{3}{}{E}$ state and zero-phonon absorption energies of the $\prescript{1}{}{A}_1$ and $\prescript{1}{}{E}$ states are from Ref.~\cite{davies1976optical, rogers2008infrared, goldman2015isc, goldman2017erratum}. We also report theoretical results obtained using $GW$ and the Bethe-Salpeter Equation (BSE)~\cite{ma2010nvgwbse}, quantum defect embedding theory (QDET)~\cite{nan2022dct}, results obtained from the constrained random-phase approximation solved by configuration interaction (CI-CRPA)~\cite{bockstedte2018ab}, and quantum chemistry results for clusters from complete active space self-consistent field (CASSCF)~\cite{bhandari2021casscf} calculations.
\textbf{b} Contribution of Slater determinants of single excitation ($|e_x\overline{e}_x\rangle$, $|e_x\overline{e}_y\rangle$, $|e_y\overline{e}_x\rangle$, $|e_y\overline{e}_y\rangle$, $|a_1\overline{e}_x\rangle$ and $|a_1\overline{e}_x\rangle$) and double excitation ($|e_x\overline{a}_1\rangle$, $|e_y\overline{a}_1\rangle$ and $|a_1\overline{a}_1\rangle$) with respect to the $\prescript{3}{}{A}_2$ ground state represented by Slater determinant $|\overline{e}_x\overline{e}_y\rangle$ to the wavefunction of the singlet states, as obtained from TDDFT and QDET~\cite{nan2022dct} calculations. Slater determinants are denoted in the hole notation, and their contributions to the total wavefunction are given in terms of the coefficients defined in equation~(\ref{eq:linearcomb}) (see text).}
\end{figure}

We computed the vertical excitation energies (VEEs) of the triplet $\prescript{3}{}{E}$ and singlet states $\prescript{1}{}{E}$ and $\prescript{1}{}{A}_1$ with respect to the $\prescript{3}{}{A}_2$ ground state using TDDFT and the semi-local functional PBE (TDDFT@PBE) and hybrid functional DDH (TDDFT@DDH), with the aim of establishing the accuracy of the chosen electronic structure methods, before proceeding with structural optimizations. Our results are shown in Fig.~\ref{fig:vee}, together with those of other calculations~\cite{ma2010nvgwbse, bockstedte2018ab, bhandari2021casscf, nan2022dct} and inferred experimental values~\cite{davies1976optical, rogers2008infrared, goldman2015isc, goldman2017erratum}. Irrespective of the functional, TDDFT correctly predicts the ordering of singlet and triplet excited states. However, at the PBE level of theory, TDDFT underestimates the energies of the $\prescript{3}{}{E}$ and the $\prescript{1}{}{A}_1$ states compared to experiment; the agreement is improved when using the hybrid functional DDH. The latter yields the energy of $\prescript{3}{}{E}$ in good accord with $GW$-BSE results~\cite{ma2010nvgwbse}, but those of the $\prescript{1}{}{E}$ and $\prescript{1}{}{A}_1$ states differ, likely due to the fact that, in contrast to $GW$-BSE, in TDDFT an approximate non-collinear spin-flip kernel is introduced to describe spin-flip excitations.

In spite of the correct ordering, the VEEs obtained at the TDDFT@DDH level of theory are an overestimate, especially for singlets, relative to the experimental values. To understand the origin of this discrepancy we compared the many-body wavefunctions obtained with TDDFT with those computed with QDET~\cite{nan2022dct}; the latter includes double and higher-order excitations from the $\prescript{3}{}{A}_2$ ground state represented by the Slater determinant $|\overline{e}_x\overline{e}_y\rangle$, that are not included in the TDDFT calculations presented here (and also in the $GW$-BSE calculations of Ref.~\cite{ma2010nvgwbse}). In QDET, the defect states are described by an effective many-body Hamiltonian diagonalized exactly by full configuration interaction (CI) and hence the many-body wavefunction contains higher-order excitations. The Hamiltonian includes the interaction of the defect and the solid where it is embedded through an effective dielectric screening. The many-body electronic wavefunctions $|\Phi_i\rangle$ are written as linear combinations of Slater determinants $|\Psi_n\rangle$:
\begin{equation}
   |\Phi_i\rangle = \sum_n c_n^i  |\Psi_n\rangle\,,
   \label{eq:linearcomb}
\end{equation}
where $|c_n^i|^2$ represents the contribution of the $n$-th Slater determinant to the $i$-th many-body electronic wavefunction. The Slater determinants with contributions to the total wavefunction larger than $1\%$ are reported in Fig.~\ref{fig:vee} and in Supplementary Table~2 for the three singlet states, for both QDET and spin-flip TDDFT calculations. Note that we use $\prescript{1}{}{A}_1^{(0)}$, $\prescript{1}{}{E}_x^{(0)}$ and $\prescript{1}{}{E}_y^{(0)}$ to denote states with $C_{3v}$ symmetry, and in Sec.~\ref{subsec:structures} we use $\prescript{1}{}{A}_1$ and $\prescript{1}{}{E}$ to denote singlet states in geometrical configurations where the $C_{3v}$ symmetry is not preserved. As shown in Fig.~\ref{fig:vee}, the major contributions to the many-body electronic states $\prescript{1}{}{A}_1^{(0)}$, $\prescript{1}{}{E}_x^{(0)}$ and $\prescript{1}{}{E}_y^{(0)}$ come from linear combinations of Slater determinants with only single excitations, which are accounted for when using spin-flip TDDFT, and yield contributions similar to QDET. However, QDET calculations show an additional, non negligible ($\sim 3 \%$) contribution to the total wavefunction coming from determinants containing double excitations that cannot be described by TDDFT: $|a_1\overline{a}_1 \rangle$, $|e_x\overline{a}_1 \rangle$, and $|e_y\overline{a}_1 \rangle$, for $\prescript{1}{}{A}_1^{(0)}$, $\prescript{1}{}{E}_x^{(0)}$ and $\prescript{1}{}{E}_y^{(0)}$, respectively.

By adding the contributions of double excitations to our spin-flip TDDFT results, using perturbation theory, we find that the energies of the $\prescript{1}{}{A}_1$ and $\prescript{1}{}{E}$ states decrease by 0.2 $\sim$ 0.3 eV, resulting in a better agreement with experiment and QDET values (see Supplementary Note~2). Hence we conclude that the absence of double excitations in the TDDFT description leads to a moderate overestimate of the energy of singlets relative to QDET results. In summary, TDDFT calculations yield results for VEEs in good (albeit not perfect) agreement with those of QDET and experiments, and account for the majority of excitations entering the many body wavefunction of the NV$^-$ center, giving us confidence that the calculation of geometries of singlet manifolds using spin-flip TDDFT is accurate.

\subsection{\label{subsec:structures} Potential energy surfaces of electronic excited states}

Having established the accuracy of TDDFT in describing VEEs, we proceed to optimize the geometry of the system in each excited state using TDDFT forces acting on nuclei. The PESs of singlets are computed by carrying out calculations on two specific geometrical paths, described by collective variables (CVs) defined below. We then define an effective Hamiltonian for ionic and electronic degrees of freedom, including electron-phonon interaction, and we investigate the non-adiabatic coupling between many-body electronic states and lattice vibrations.

We start by describing the optimized geometrical configurations of excited electronic states, quantified in terms of mass-weighted atomic displacements and Franck-Condon shifts (see Supplementary Note~3). We find that the optimized geometry of the triplet excited state $\prescript{3}{}{E}$ exhibits a significant displacement of $\sim$0.6 amu$^{0.5}$ \AA\ and a Franck-Condon shift of $\sim$200 meV, relative to the geometry of the ground state. These  results obtained with TDDFT forces are consistent with our previous study, where geometry optimization of the triplet excited state was obtained with $\Delta$SCF, and results were validated against photoluminescence (PL) measurements~\cite{jin2021pl}. The singlet states cannot be simulated with $\Delta$SCF. Hence, we optimize their geometry using forces computed with spin-flip TDDFT and a plane-wave basis set. The two singlet states have rather different optimized configurations: that of the $\prescript{1}{}{A}_1$ state is similar to the optimized geometry of the ground state (with a negligible atomic displacement of $\sim$0.1 amu$^{0.5}$ \AA\ and a Franck-Condon shift of 17 meV), while the $\prescript{1}{}{E}$ state exhibits a displacement of $\sim$0.4 amu$^{0.5}$ \AA\ and a Franck-Condon shift of 60$\sim$100 meV.

\begin{figure}[t]
\includegraphics[width=9cm]{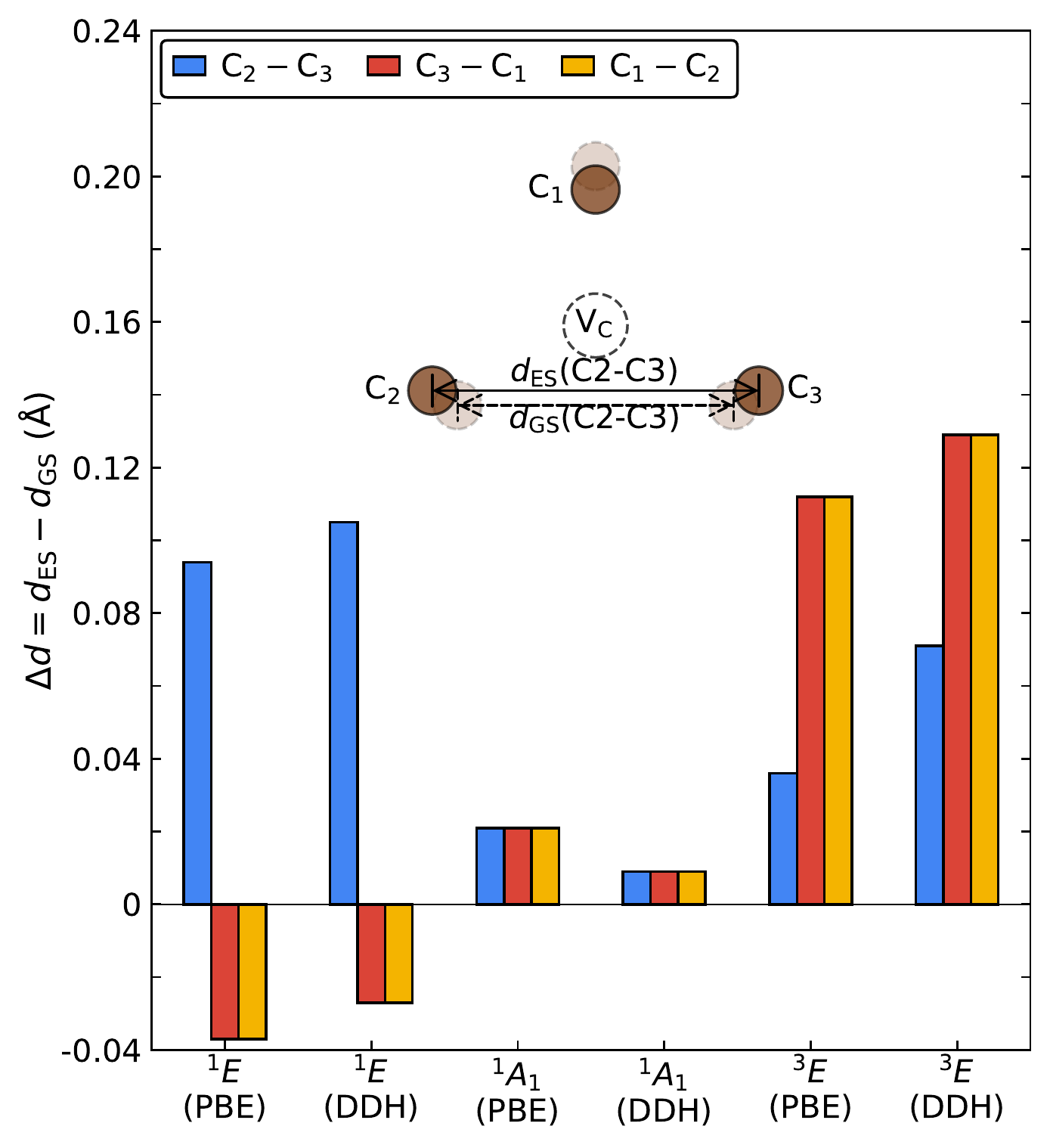}
\caption{\label{fig:dis} \textbf{Geometrical configurations of many-body states of the NV$^-$ center in diamond.} Differences of the distances between the three carbon atoms (C$_1$, C$_2$ and C$_3$) around the vacancy site (V$_{\text{C}}$), as obtained in the excited states (ES) and ground state (GS): $\Delta d = d_{\text{ES}} - d_{\text{GS}}$. The differences are reported for the $\prescript{1}{}{E}$, $\prescript{1}{}{A}_1$ and $\prescript{3}{}{E}$ excited states (ESs) and are computed using TDDFT with PBE or DDH functionals. Note that $\Delta d$(C$_1-$C$_2$), $\Delta d$(C$_2-$C$_3$) and $\Delta d$(C$_3-$C$_1$) for the $\prescript{3}{}{E}$ state differ, due to the coupling of the electronic state to both $a_1$ and $e$ type phonon modes (see text). For the $\prescript{1}{}{A}_1$ state, $\Delta d$(C$_1-$C$_2$), $\Delta d$(C$_2-$C$_3$) and $\Delta d$(C$_3-$C$_1$) are instead all equal within $\sim$ 0.02 \AA, implying that in this case the equilibrium geometry preserves the $C_{3v}$ symmetry and is close to that of the $\prescript{3}{}{A}_2$ ground state. The differences $\Delta d$(C$_1-$C$_2$), $\Delta d$(C$_2-$C$_3$) and $\Delta d$(C$_3-$C$_1$) of the $\prescript{1}{}{E}$ state differ, due to a significant coupling with $e$ type phonon modes, leading to symmetry breaking.}
\end{figure}

We then computed the variation of the distances ($\Delta d$) of the three carbon atoms close to $\text{V}_\text{C}$ in the excited states ($d_{\text{ES}}$), relative to the ground state ($d_{\text{GS}}$); these are shown in Fig.~\ref{fig:dis}. We find an asymmetric displacement pattern for the $\prescript{1}{}{E}$ singlet, suggesting the existence of three equivalent equilibrium geometries, compatible with the $C_{3v}$ symmetry of the defect, which we characterized in terms of two CVs, $Q_\alpha$ and $Q_\beta$. $Q_\beta$ defines a direction connecting two of the three geometrical configurations, and $Q_\alpha$ is perpendicular to $Q_\beta$. The three geometrical configurations form an equilateral triangle on the plane defined by $Q_\alpha$ and $Q_\beta$. The minimum of the $\prescript{1}{}{A}_1$ singlet PES on the plane of $Q_\alpha$ and $Q_\beta$ is located at the center of the triangle (defined by $Q_\alpha=0$, $Q_\beta=0$), and is very close to the actual minimum of the $\prescript{1}{}{A}_1$ singlet with a negligible displacement of 0.08 amu$^{1/2}$ \AA. Using the CVs $Q_\alpha$ and $Q_\beta$ we computed the total energies of the singlet many-body states along two paths, using TDDFT@PBE: path 1, parallel to $Q_\alpha$, with $Q_\beta = 0$, which connects one of the local minima and the center of the triangle; path 2, parallel to $Q_\beta$, with $Q_\alpha = 0$, and crossing the triangle center (see Supplementary Fig.~1). For values of $Q_\alpha$ and $Q_\beta$ different from zero, and along both paths 1 and 2, we find that the wavefunctions of the $\prescript{1}{}{A}_1$ and $\prescript{1}{}{E}$ singlets, as computed using TDDFT, are linear combinations of the states with $C_{3v}$ symmetry previously identified as $\prescript{1}{}{A}_1^{(0)}$, $\prescript{1}{}{E}_x^{(0)}$ and $\prescript{1}{}{E}_y^{(0)}$.  For the $\prescript{1}{}{A}_1$ singlet, the wavefunction is given by a linear combination of  the $\prescript{1}{}{A}_1^{(0)}$ component, mixed with a small amount ($<$10$\%$) of the $\prescript{1}{}{E}_x^{(0)}$ component along path 1 (or $\prescript{1}{}{E}_y^{(0)}$ component along path 2). The magnitude of the mixing between states with $C_{3v}$ symmetry increases as the absolute value of $Q_\alpha$ and $Q_\beta$ increases. While the wavefunction of the $\prescript{1}{}{E}$ singlet on path 1 can still be approximately identified as the so called ``pure'' state $\prescript{1}{}{E}_x^{(0)}$ or $\prescript{1}{}{E}_y^{(0)}$, on path 2, the wavefunction is given by a linear combination with approximately equal weights of the $\prescript{1}{}{E}_x^{(0)}$ and $\prescript{1}{}{E}_y^{(0)}$ components. The mixing of components found in our calculations points at the non-adiabatic coupling occurring in the system, which requires further analysis, as we discuss next.

\begin{figure}[t]
\vspace{-40pt}
\includegraphics[width=16cm]{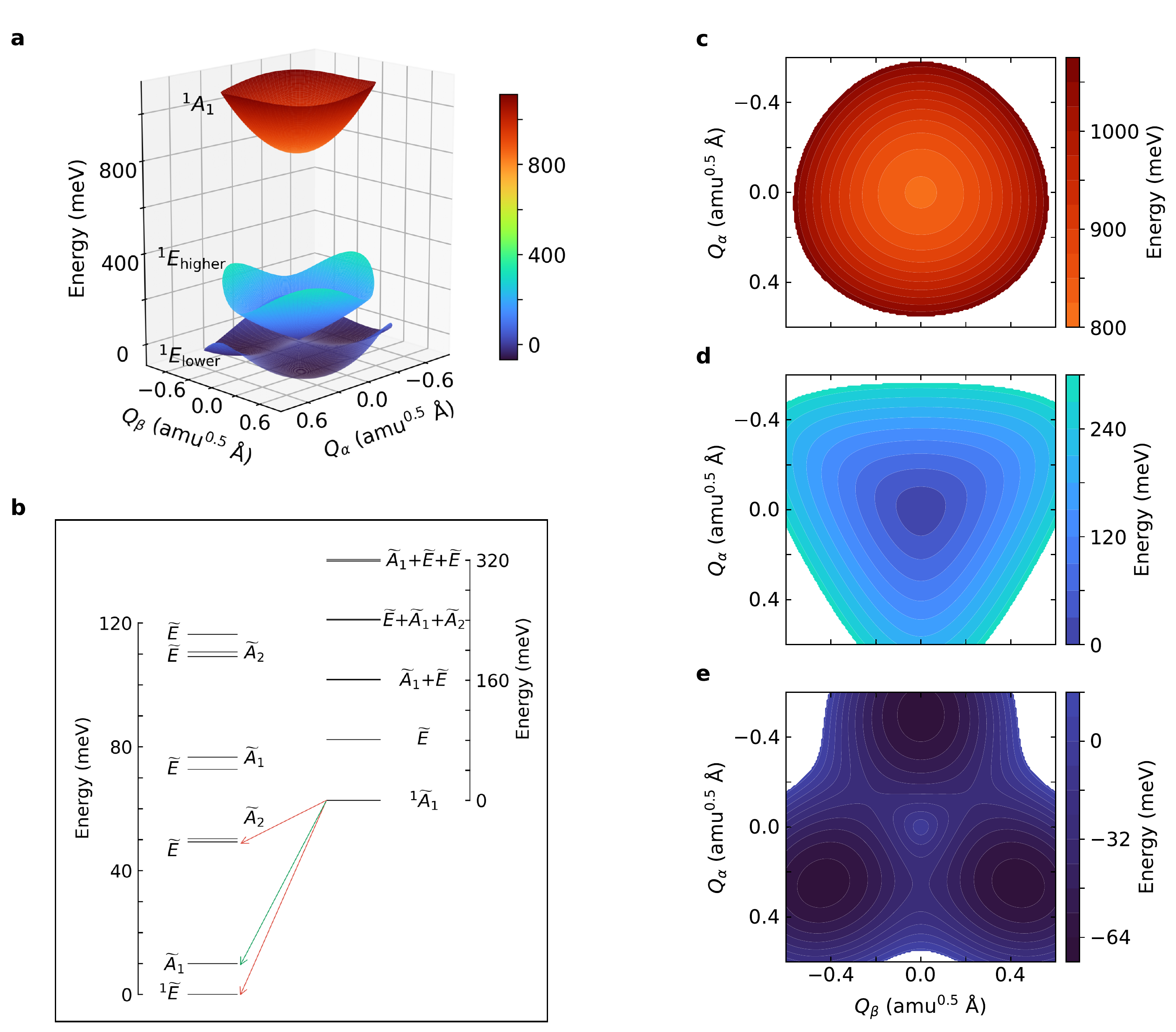}
\vspace{-20pt}
\caption{\label{fig:pes}\textbf{Potential energy surfaces (PESs) and vibronic energy levels of the many-body electronic states of the NV$^-$ center in diamond.} \textbf{a} Adiabatic PESs of the lower and higher branches of the $\prescript{1}{}{E}$ and $\prescript{1}{}{A}_1$ states. The $Q_\alpha$, $Q_\beta$ configuration coordinates (see text) represent the collective motion of effective phonon modes with $e$ symmetry. Contour plots of the PESs are shown in  \textbf{c-e}. The PES of the $\prescript{1}{}{E}$ lower branch (\textbf{e}) has the ``tricorn Mexican hat'' shape with three minima and three saddle points, and is connected to the higher branch (\textbf{d}) through a cusp. The PES of the $\prescript{1}{}{A}_1$ singlet (\textbf{c}) slightly deviates from a perfect two-dimensional paraboloid. \textbf{b} The vibronic levels of the $\prescript{1}{}{E}$ (left) and $\prescript{1}{}{A}_1$ (right) states, whose vibronic ground states are labelled as $\prescript{1}{}{\widetilde{E}}$ and $\prescript{1}{}{\widetilde{A}}_1$, respectively. The energy differences of vibronic levels of the $\prescript{1}{}{A}_1$ from bottom to top are found to be 80.8 meV, 79.7 meV, 78.8 meV, 77.3 meV, respectively. The selection rules for the photoluminescence (PL) are indicated as arrows: red arrows represent the optically active $\prescript{1}{}{\widetilde{A}}_1 \to \prescript{1}{}{\widetilde{E}}$ transition resulting in the zero-phonon line (ZPL) and the $\prescript{1}{}{\widetilde{A}}_1 \to \widetilde{E}$ transition resulting in the phonon side band shifted by 49.3 meV from the ZPL; the green arrow represents the $\prescript{1}{}{\widetilde{A}}_1 \to \widetilde{A}_1$ transition, shifted by 10.0 meV from the ZPL, and can be activated by uniaxial stress. In the plot, the values of the ZPL and side bands are not given on the same energy scale for clarity.}
\end{figure}

To analyze in detail the PESs of the singlet states, we define an effective Hamiltonian that includes electron-phonon (non-adiabatic) coupling~\cite{bersuker2006jt, thiering2018nvisc}, and where the nuclei are represented in terms of the CVs defined above, and the electrons in the basis of the three singlet states $\prescript{1}{}{A}_1^{(0)}$, $\prescript{1}{}{E}_x^{(0)}$ and $\prescript{1}{}{E}_y^{(0)}$ at $Q_\alpha =0$ and $Q_\beta = 0$:
\begin{equation}
\label{eq:heff}
    \hat{H} = \hat{H}_{e} + \hat{H}_{ph} + \hat{H}_{e-ph}.
\end{equation}
Here $\hat{H}_e = \sum_i E_i \hat{c}_i^{\dagger} \hat{c}_i$ is the electronic Hamiltonian, and $\hat{c}^{\dagger}_i$ ($\hat{c}_i$) is the creation (annihilation) operator of the $i$-th many-body electronic state with $E_i = \left(\Lambda, 0, 0\right)$ for $|\Phi_i\rangle = \left( |\prescript{1}{}{A}_1^{(0)}\rangle, |\prescript{1}{}{E}_x^{(0)}\rangle, |\prescript{1}{}{E}_y^{(0)}\rangle\right)$; $\Lambda=821 \,\text{meV}$ is the energy gap between the $\prescript{1}{}{A}_1^{(0)}$ and degenerate $\prescript{1}{}{E}_x^{(0)}$ and $\prescript{1}{}{E}_y^{(0)}$ electronic states obtained with TDDFT@PBE. $\hat{H}_{ph} = \sum_{\lambda = \alpha, \beta} \hbar\omega_e \left( \hat{b}_{\lambda}^{\dagger} \hat{b}_{\lambda} + \frac{1}{2} \right)$  is the Hamiltonian of the 2D harmonic oscillator written in terms of $Q_\alpha$ and $Q_\beta$, with an effective phonon energy of $\hbar\omega_e$, and $\hat{b}^{\dagger}_\lambda$ ($\hat{b}_\lambda$) is the creation (annihilation) operator of phonon $\lambda$. The electron-phonon coupling term reads
\begin{equation}
    \hat{H}_{e-ph} = \sum_{ij}\sum_{\lambda=\alpha,\beta} g_{ij,\lambda} \hat{c}_i^{\dagger} \hat{c}_j \left( \hat{b}_{\lambda}^{\dagger} + \hat{b}_{\lambda} \right),
\end{equation}
where $g_{ij,\lambda}$ is the linear electron-phonon  coupling strength between electronic state $i$, $j$ and phonon mode $\lambda$. Details on our first-principles calculation of the electron-phonon coupling strength and the analysis of the Hamiltonian of equation~(\ref{eq:heff}) in terms of pseudo- and dynamical Jahn-Teller effects are given in Supplementary Note~4.

To obtain the adiabatic PESs of the singlet states we write $\hat{b}_{\lambda} = \sqrt{\frac{\omega_e}{2\hbar}} \left(\hat{Q}_{\lambda} + \frac{i}{\omega_e} \hat{\Pi}_{\lambda} \right)$, where $\hat{\Pi}_{\lambda}$ is the momentum operator. Treating $Q_{\lambda}$ and $\Pi_{\lambda}$ as classical coordinates allows us to separate the kinetic and potential energy terms in the Hamiltonian, and hence to obtain the adiabatic PESs, which are displayed in Fig.~\ref{fig:pes} \textbf{c-e}. We obtained the parameters of the Hamiltonian, including the effective phonon energy $\hbar\omega_e = 63 \,\text{meV}$ and the electron-phonon coupling strength $g_{ij,\lambda}$, by fitting the PESs obtained with the Hamiltonian equation~(\ref{eq:heff}) to  our first-principles calculations, without introducing any empirical parameters (see Supplementary Fig.~2). The lower branch of the PES of the $\prescript{1}{}{E}$ singlet exhibits a ``tricorn Mexican hat'' shape with three minima and three saddle points, and is connected to the higher branch through a cusp. The PES of the $\prescript{1}{}{A}_1$ singlet slightly deviates from a perfect two-dimensional paraboloid, and the anharmonicity is most apparent along the path connecting its minimum to the minima on the lower branch of the $\prescript{1}{}{E}$ state PES.

By solving the effective Hamiltonian equation~(\ref{eq:heff}) considering quantized vibrations, instead of classical coordinates, we obtain the vibronic levels of the two singlet states, as shown in Fig.~\ref{fig:pes} \textbf{b}. We find that the vibronic levels with major electronic contribution from the $\prescript{1}{}{A}$ singlet state are well approximated by harmonic vibrational levels, being almost equidistant with an energy gap of $\sim80$ meV. The energy gap is $17$ meV higher than the energy of the effective phonon defined in equation~(\ref{eq:heff}), as a result of the non-adiabatic coupling. Non-adiabatic coupling also results in noticeable anharmonicity: the energy difference between adjacent vibronic levels with major contribution coming from $\prescript{1}{}{A}_1$ decreases as the quantum number increases. On the other hand, the vibronic levels with major electronic contribution from the $\prescript{1}{}{E}$ singlet state are substantially different from those of a quantum harmonic oscillator. Our calculations identify an $\prescript{1}{}{\widetilde{A}}_1$ state 10 meV above the vibronic ground state ($\prescript{1}{}{\widetilde{E}}$), which likely corresponds to the state detected experimentally at about 14 to 16 meV~\cite{manson2010optically, rogers2015singlet,acosta2010optical, robledo2011spin}, and discussed in Ref.~\cite{thiering2018nvisc}. Such state is not accessible under equilibrium conditions but can be reached when the crystal is under uniaxial stress. We also find degenerate $\prescript{1}{}{\widetilde{E}}$ vibronic levels at 49.3 meV above the vibronic ground state; the transition into these states might be the origin of the phonon side band at 42.6 meV observed in the low-temperature experimental PL spectrum of the $\prescript{1}{}{A}_1 \to \prescript{1}{}{E}$ transition~\cite{rogers2008infrared}.

Finally we note that, unlike the $\prescript{1}{}{E} \to \prescript{1}{}{A_1}$ absorption line shape, the calculation of the $\prescript{1}{}{A}_1 \to \prescript{1}{}{E}$ PL line shape would require an evaluation of all the phonon modes of the $\prescript{1}{}{E}$ state, whose PES is strongly anharmonic, as well as an explicit treatment of the non-adiabatic coupling including all phonon modes~\cite{razinkovas2021nvpl}. Although in principle possible, these calculations are beyond the scope of the present work.

\subsection{\label{subsec:phonon} Optical spectra}

We now turn to the discussion of our calculations of the vibrationally resolved absorption spectrum for the transition between singlet states, which we compare with experiments and with the PL spectrum for the transition between triplet states. 

\begin{figure}[t]
\includegraphics[width=16cm]{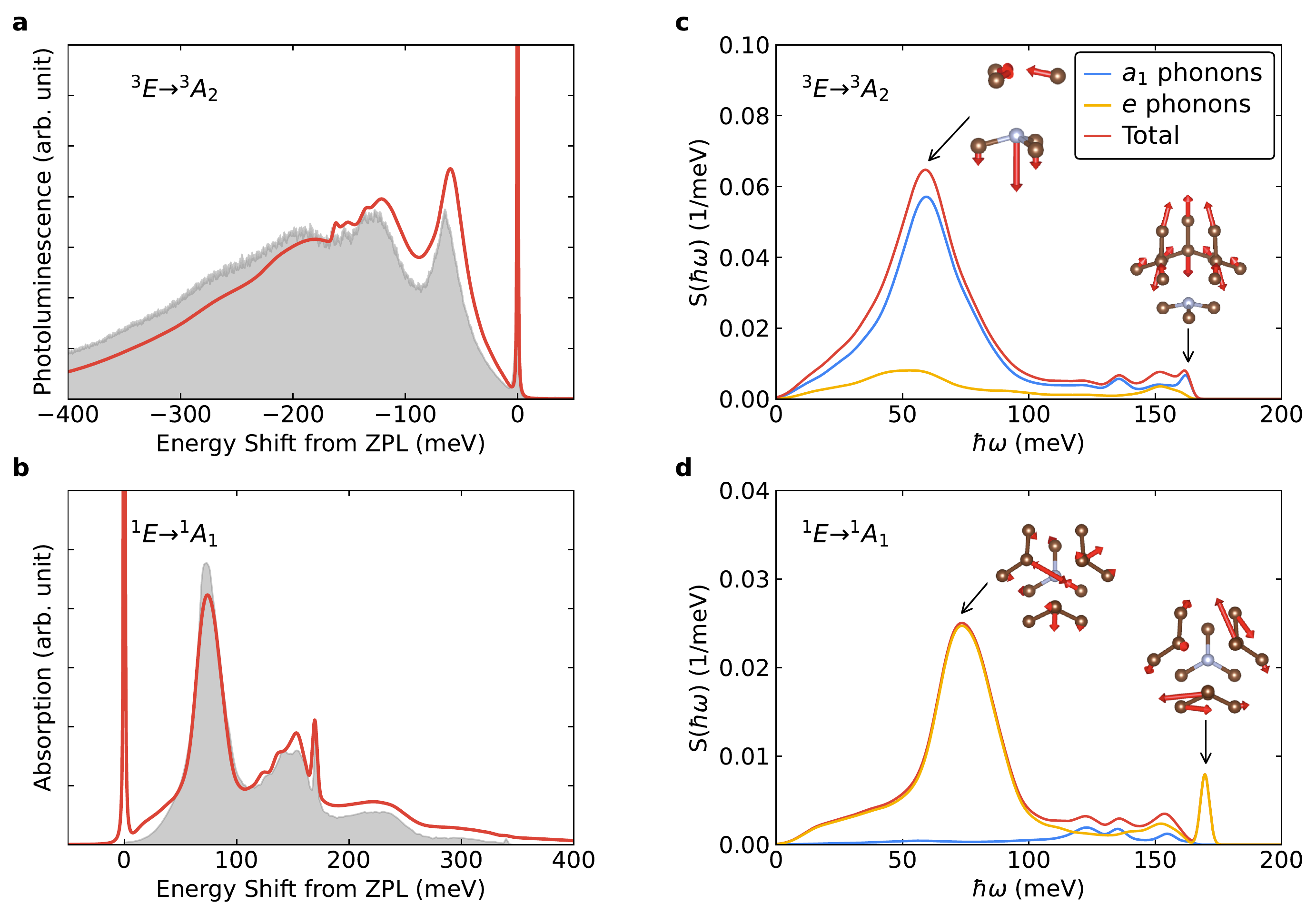}
\caption{\label{fig:phonon} \textbf{Optical spectra and spectral densities.} \textbf{a} Photoluminescence (PL) line shapes of the $\prescript{3}{}{E} \to \prescript{3}{}{A}_2$ transition and \textbf{b} absorption line shapes of the $\prescript{1}{}{E} \to \prescript{1}{}{A_1}$ transition. Red lines are theoretical results while the gray area represents experimental spectra from Ref.~\cite{kehayias2013nvexp, alkauskas2014luminescence}. 
Spectral densities $S(\hbar\omega)$ of the $\prescript{3}{}{E} \to \prescript{3}{}{A}_2$ (\textbf{c}) and the $\prescript{1}{}{E} \to \prescript{1}{}{A_1}$ transitions (\textbf{d}). Contributions from $a_1$ and $e$ type phonon modes are shown as blue and yellow lines, respectively. The quasi-local (local) $a_1$ mode at 60 meV (162 meV) of the $\prescript{3}{}{A}_2$ state that strongly couples with the $\prescript{3}{}{E} \to \prescript{3}{}{A}_2$ transition is shown in the inset of \textbf{a}. The quasi-local (local) $e$ mode at 73 meV (170 meV) of the $\prescript{1}{}{A}_1$ state that strongly couples with the $\prescript{1}{}{E} \to \prescript{1}{}{A}_1$ transition is shown in the inset of \textbf{c}. Results reported here are based on phonons computed at the PBE level of theory and optimized geometries computed at the DDH level of theory, and are extrapolated to the dilute limit, approximated by a $(12 \times 12 \times 12)$ supercell with 13824 atomic sites. A comparison of results obtained using different functionals is given in Supplementary Note~5.}
\end{figure}

Having computed the forces acting on nuclei with spin-flip TDDFT and all phonon modes in the $\prescript{1}{}{A}_1$ state, we calculated the vibrationally resolved absorption spectrum of the transition between the $\prescript{1}{}{E}$ and $\prescript{1}{}{A}_1$ singlets using the HR theory. At T $\sim$ 0 K, transitions occur from the lowest vibronic level of the $\prescript{1}{}{E}$ state whose vibronic wavefunction is localized in the local minimum of the PES, into vibronic levels of the $\prescript{1}{}{A}_1$ singlet state; these levels are all well approximated by harmonic vibrational levels; hence the use of the HR theory is justified. 

Our results are compared with experiment~\cite{kehayias2013nvexp} in Fig.~\ref{fig:phonon}. The agreement is excellent (see Supplementary Note~5 for a comparison of results obtained using different functionals), and we successfully predict for the first time the main peak at 73 meV and the sharp peak at 170 meV. Note that the energy of the main peak is 7 meV smaller than the distance between vibronic levels of the $\prescript{1}{}{A}_1$ state obtained from the effective Hamiltonian equation~(\ref{eq:heff}), pointing at the importance of including all phonon modes in the calculation of optical spectra. The level of agreement obtained here indicates that our first-principles calculations based on spin-flip TDDFT correctly describe the atomic geometry relaxation and vibrational properties of the singlet states. Such properties are not accessible in $\Delta$SCF and hence their calculations require the implementation of TDDFT forces. In addition, we emphasize the importance of including the anharmonicity of the PES of the $\prescript{1}{}{A}_1$ singlet in the calculation of the HR factors and spectral functions (see Supplementary Note~7).

Note that the phonon side band of the absorption line shape for the $\prescript{1}{}{E} \to \prescript{1}{}{A}$ transition decays much faster compared with that of the PL spectrum for the $\prescript{3}{}{E} \to \prescript{3}{}{A}_2$ transition (shown in Fig.~\ref{fig:phonon} \textbf{a} for comparison). Indeed, the computed Debye-Waller factor (the ratio of the zero-phonon line (ZPL) relative to the entire line shape) of the $\prescript{1}{}{E} \to \prescript{1}{}{A}$ absorption line shape is 34$\%$, in good agreement with the inferred experimental value of $\sim$ 40\%~\cite{kehayias2013nvexp}, and is 10 times larger than that of the $\prescript{3}{}{E} \to \prescript{3}{}{A}_2$ PL line shape. The large Debye-Waller factor suggests that the ZPL is more absorptive than the phonon side band and hence better suited for infrared-absorption-based magnetometry measurements than the phonon side band wavelengths~\cite{kehayias2013nvexp}.

It is interesting to analyze the main differences between singlet absorption and triplet PL spectra in terms of the spectral density of the electron-phonon coupling in the two cases, $S(\hbar\omega)$, as shown in Fig.~\ref{fig:phonon} \textbf{c} and \textbf{d}. The main contribution to the $S(\hbar\omega)$ of the $\prescript{1}{}{E} \to \prescript{1}{}{A}_1$ transition comes from the coupling of the electronic states with $e$ type phonon modes; instead the main contribution in the case of the $\prescript{3}{}{E} \to \prescript{3}{}{A}_2$ transition originates from the coupling with $a_1$ type phonon modes. In more detail, we find that $S(\hbar\omega)$ of the $\prescript{1}{}{E} \to \prescript{1}{}{A}_1$ transition exhibits a broad peak at 73 meV and a sharp peak at 170 meV, resulting from the coupling of the electronic states with a quasi-local and a local $e$ type phonon mode, displayed in the inset of Fig.~\ref{fig:phonon} \textbf{d}. The 170 meV $e$ type local phonon mode exists only in the $\prescript{1}{}{A}_1$ state and has an energy higher than that of the optical phonons of diamond. The $S(\hbar\omega)$ of the $\prescript{1}{}{E} \to \prescript{1}{}{A_1}$ transition is generally shifted to higher energy relative to that of the $\prescript{3}{}{E} \to \prescript{3}{}{A_2}$ transition, originating from an increase of the energy of the phonons of the $\prescript{1}{}{A}_1$ state compared with those of the $\prescript{3}{}{A}_2$ state. Previous work suggested that such an increase of phonon energies might be caused by the contribution of the double excitation configurations $|a_1\overline{a}_1\rangle$ in the wavefunction of the $\prescript{1}{}{A}_1$ state~\cite{kehayias2013nvexp}. However, our work suggests that the non-adiabatic coupling of the $\prescript{1}{}{A}_1$ and $\prescript{1}{}{E}$ singlet states is more likely responsible for the increase in phonon energies. A detailed comparison of the phonon modes of the $\prescript{1}{}{A}_1$ and $\prescript{3}{}{A}_2$ states can be found in Supplementary Note~6.

\section{\label{sec:discussion}Discussion}
In summary, we studied the many-body electronic states of the NV$^-$ center in diamond, including singlet states, using first-principles calculations based on TDDFT with semi-local and hybrid functionals and we computed vibrationally resolved optical spectra. Our work represents the first application of TDDFT with analytical forces to the prediction of optical spectra of spin-defects in solids. We found that TDDFT provides an accurate description of both the electronic structure and atomic geometries of the many-body electronic states. In particular, TDDFT predicts the same energy ordering as experiment and the correct characteristics of the many-body electronic states, similar to those obtained using higher level methods, although the neglect of double excitations results in a slight overestimate of excitation energies relative to experiments. The computed vibrationally resolved absorption spectrum of the $\prescript{1}{}{E} \to \prescript{1}{}{A_1}$ transition is in excellent agreement with experiment, thanks to an accurate description of the atomic geometries and phonons of the singlet states obtained in our work. Our results show the key role played by non-adiabatic coupling in determining optical transitions. For example, we found that the equilibrium  geometry of the $\prescript{1}{}{A}_1$ state is similar to that of the $\prescript{3}{}{A}_2$ ground state; however, the $e$ type phonons of the former have significant higher energy than those of the ground state, due to the non-adiabatic coupling of the former with the $\prescript{1}{}{E}$ states. Such coupling is also responsible for the anharmonicity of the $\prescript{1}{}{A}_1$ state PES which should be taken into account in obtaining absorption spectra in quantitative agreement with experiment. Interestingly, solving the effective Hamiltonian for the non-adiabatic coupling yields optically forbidden $\widetilde{A}_1$ and optically allowed $\widetilde{E}$ vibronic levels above the $\prescript{1}{}{\widetilde{E}}$ ground vibronic state, consistent with PL measurements. Our study provides first principles predictions of the basic properties of the NV$^-$ center in diamond, which are important for a comprehensive understanding of the optical spin-polarization cycle of this defect, and hence of its functionalities for quantum technology applications. In particular, the techniques presented here enable the modeling, from first principles, of the phonon side band of the optical absorption process between singlet states, which has been used for infrared absorption-based magnetometry. The strategy applied here to the NV$^-$ center in diamond is general and paves the way to the study of shelving states and optical spectra in other spin defects and materials.

\section{\label{sec:methods}Methods}

\subsection{\label{sec:dft}Electronic structure calculations}
The ground state electronic structure of the NV$^-$ center in diamond was obtained using DFT and the planewave pseudopotential method, as implemented in the Quantum Espresso package~\cite{QE-2009, QE-2017, QE-exascale}. We used SG15 ONCV norm-conserving pseudopotentials~\cite{ONCV_1, SCHLIPF201536_ONCV_2} and the semi-local functional by Perdew, Burke, and Ernzerhof (PBE)~\cite{PBE} and the dielectric dependent hybrid (DDH) functional~\cite{skone2014ddh}. The fraction of exact exchange used in the DDH functional is the inverse of the macroscopic dielectric constant of the system as reported in Ref.~\cite{skone2014ddh,hosung2017qubit}. The planewave energy cutoff was set to 85 Ry when using the PBE functional, and to 60 Ry for the  DDH functional. We used a $(3\times 3\times 3)$ supercell containing 216 atomic sites for the NV$^-$ center in diamond, with the lattice constant optimized for each functional~\cite{jin2021pl}. The convergence of our results for VEEs with respect to the supercell size is reported in Supplementary Note~1. The Brillouin zone of the supercell was sampled with the $\Gamma$ point.

Excited states were computed using the TDDFT method within the Tamm-Dancoff approximation. We obtained the energies and eigenvectors of low-lying excited states by iteratively diagonalization of the linearized Liouville operator, as implemented in the WEST code~\cite{govoni2015west, nguyen2019finite}. An approximated non-collinear kernel was included in the spin-flip TDDFT calculations~\cite{wang2004sftddft,wang2005sftddft,li2012sftddft}. Analytical forces on nuclei in TDDFT were evaluated using the Lagrangian formulation by H\"{u}tter~\cite{hutter2003excited}. The equilibrium atomic geometries of excited states were obtained by minimizing the nuclear forces below the threshold of 0.01 eV/\AA.

\subsection{\label{sec:phonon}Phonon calculations}
Phonon modes of the NV$^-$ center were computed using the frozen phonon approach, with configurations generated with the PHONOPY package~\cite{phonopy} and a displacement of 0.01 \AA{} from equilibrium geometries of the $\prescript{3}{}{A}_2$ and $\prescript{1}{}{A}_1$ states, respectively. To compute the phonon modes of the the $\prescript{3}{}{A}_2$ state, DFT self-consistent calculations were conducted at displaced configurations. As for the phonon modes of the $\prescript{1}{}{A}_1$ state, a DFT self-consistent calculation and an additional TDDFT excited state calculation were performed at each displaced configuration. Phonon calculations were performed only with the PBE functional due to the high computational cost of hybrid DFT calculations. We estimated hybrid-DFT phonons by using a scaling factor~\cite{jin2021pl}. Phonon modes are extrapolated to the dilute limit, approximated by a $(12 \times 12 \times 12)$ supercell cell with 13824 atomic sites, using the force constant matrix embedding approach proposed by Alkauskas et al.~\cite{alkauskas2014luminescence, razinkovas2021nvpl}.

\subsection{\label{sec:pl} Huang-Rhys factors and spectral functions}
We write the absorption line shape as~\cite{razinkovas2021photoionization}
\begin{equation}
    \sigma_{\text{abs}}(\hbar\omega, T) \propto (\hbar\omega) A_{\text{abs}} (\hbar\omega - E_{\text{ZPL}}, T),
\end{equation}
where $E_{\text{ZPL}}$ is the energy of the zero-phonon line, and $\hbar\omega$ is the energy of the absorbed photon. $T$ is the temperature. To be consistent with experiment~\cite{kehayias2013nvexp}, $T = 10$ K was used in the calculation of absorption line shape.  The absorption spectral function is computed using the generating function approach~\cite{kubo1955generatingfxn, lax1952generatingfxn, alkauskas2014luminescence}
\begin{equation}
    A_{\text{abs}} (\hbar\omega, T) = \dfrac{1}{2\pi} \int_{-\infty}^{\infty} e^{i\omega t}  G_{\text{abs}}(t, T) e^{- \frac{\lambda|t|}{\hbar}} \mathrm{d} t,
\end{equation}
where $\lambda = 0.1$ meV was used in our calculation to account for the broadening of the line shape. The generating function is written as
\begin{equation}
\begin{aligned}
    G_{\text{abs}}(t,T) &= \exp\bigg[ \int_{-\infty}^{\infty} S(\hbar\omega) e^{-i\omega t} \mathrm{d} (\hbar\omega) - \sum_k S_k \\
    &+  \int_{-\infty}^{\infty} C(\hbar\omega, T) e^{-i\omega t} \mathrm{d} (\hbar\omega) + \int_{-\infty}^{\infty} C(\hbar\omega, T) e^{i\omega t} \mathrm{d} (\hbar\omega) - 2\sum_k \overline{n}_k(T)S_k\bigg],
\end{aligned}
\end{equation}
where $\overline{n}_k (T)$ is the average occupation number of the $k$th phonon mode. $S(\hbar\omega)$ and $C(\hbar\omega, T)$ are the spectral densities of electron-phonon coupling,
\begin{equation}
\begin{aligned}
    S(\hbar\omega) &= \sum_{k} S_k \delta(\hbar\omega - \hbar\omega_k), \quad 
    C(\hbar\omega, T) &= \sum_{k} \overline{n}_k(T) S_k \delta(\hbar\omega - \hbar\omega_k).
\end{aligned}
\end{equation}
In actual calculations, the $\delta$ functions are replaced by Gaussian functions, and the broadening $\sigma_k$ is varied linearly from 6 to 2 meV with the phonon energy, to account for the continuum of phonon modes participating in the optical transition. The HR factor $S_k$ is computed as
\begin{equation}
    S_k = \frac{\omega_k \Delta Q_k^2}{2\hbar}.
\end{equation}
where $\Delta Q_k$ is the mass-weighted displacement along the $k$th mode, evaluated as
\begin{equation}
    \Delta Q_k = \frac{1}{\omega_k^2} \sum_{\alpha=1}^N \sum_{i=x,y,z} \frac{\mathbf{F}_{\alpha i}}{\sqrt{M_\alpha}} \mathbf{e}_{k,\alpha i}.
\end{equation}
Here $\mathbf{e}_{k,\alpha i}$ is the eigenvector of the $k$th phonon mode, $M_{\alpha}$ is the mass of the $\alpha$th atom. For the $\prescript{1}{}{E} \to \prescript{1}{}{A}_1$ absorption, $\mathbf{F}$ represents the forces of the $\prescript{1}{}{A}_1$ state evaluated at the equilibrium geometry of the $\prescript{1}{}{E}$ state. $\omega_k$ ($\mathbf{e}_k$) is the frequency (eigenvector) of the phonons of the $\prescript{1}{}{A}_1$ state.

Similarly, the PL line shape of the $\prescript{3}{}{A}_2 \to \prescript{3}{}{E}$ transition can be computed as~\cite{razinkovas2021photoionization}
\begin{equation}
    I(\hbar\omega, T) \propto (\hbar\omega)^3 A_{\text{emi}}(E_{\text{ZPL}} - \hbar\omega, T).
\end{equation}
Here the emission spectral function is calculated using the generating function built on HR factors computed with forces of the $\prescript{3}{}{A}_2$ state, evaluated at the equilibrium structure of the $\prescript{3}{}{E}$ state and with the phonons of the $\prescript{3}{}{A}_2$ state. To be consistent with experiment~\cite{alkauskas2014luminescence}, $T = 8$ K was used in the calculation of the PL line shape.

\section{Code \& data availability}
The TDDFT calculations and analytical nuclear forces are  implemented in the open source code \texttt{WEST} (west-code.org/).

Data that support the findings of this study will be available through the \texttt{Qresp}~\cite{govoni2019qresp} curator at \href{https://paperstack.uchicago.edu/explorer}{https://paperstack.uchicago.edu/explorer}.

\section{Acknowledgement}
We thank Dr. He Ma, Nan Sheng and Dr. Christian Vorwerk for fruitful discussions. This work was supported by the computational materials science center Midwest Integrated Center for Computational Materials (MICCoM) for the implementation of spin-flip time-dependent density functional theory in WEST, and by AFOSR Grant No. FA9550-19-1-0358 for the application of the code to study absorption spectra in diamond. MICCoM is part of the Computational Materials Sciences Program funded by the U.S. Department of Energy, Office of Science, Basic Energy Sciences, Materials Sciences, and Engineering Division through the Argonne National Laboratory, under Contract No. DE-AC02-06CH11357. This research used  resources of the National Energy Research Scientific Computing Center (NERSC), a DOE Office of Science User Facility supported by the Office of Science of the U.S. Department of Energy under Contract No. DE-AC02-05CH11231, and resources of the University of Chicago Research Computing Center.

\section{Author Contributions}
Y.J., M.G., and G.G. designed the research. Y.J. implemented the time-dependent density functional theory and analytical nuclear forces and performed calculations, with supervision by M.G. and G.G. All authors wrote the manuscript.

\section{Competing Interests}
The authors declare no competing interests.

\providecommand{\noopsort}[1]{}\providecommand{\singleletter}[1]{#1} %

\bibliographystyle{unsrt}

\end{document}


\title{Supplementary Information: Vibrationally resolved optical excitations of the nitrogen-vacancy center in diamond
}

\author{Yu Jin}
\affiliation{Department of Chemistry, University of Chicago, Chicago, Illinois 60637, United States}
\author{Marco Govoni}
\email{mgovoni@anl.gov}
\affiliation{Pritzker School of Molecular Engineering, University of Chicago, Chicago, Illinois 60637, United States}
\affiliation{Materials Science Division and Center for Molecular Engineering, Argonne National Laboratory, Lemont, Illinois 60439, United States}
\author{Giulia Galli}
\email{gagalli@uchicago.edu}
\affiliation{Department of Chemistry, University of Chicago, Chicago, Illinois 60637, United States}
\affiliation{Pritzker School of Molecular Engineering, University of Chicago, Chicago, Illinois 60637, United States}
\affiliation{Materials Science Division and Center for Molecular Engineering, Argonne National Laboratory, Lemont, Illinois 60439, United States}

\date{\today}

\maketitle

\tableofcontents
\newpage

\section*{\label{s-sec:size} Supplementary Note 1: Supercell size convergence}
To check convergence with respect to the size of the supercell, we performed TDDFT calculations at the PBE level of theory for supercells with up to 512 atomic sites, corresponding to a supercell with $(4\times 4 \times 4)$ repetitions of the unit cell. Vertical excitation energies (VEEs) of many-body excited states are summarized in Supplementary Table~\ref{s-tab:size}. We found that the difference in the VEEs calculated with the $(3 \times 3 \times 3)$ and the $(4 \times 4 \times 4)$ supercells is less than 0.02 eV, indicating that the $(3 \times 3 \times 3)$ supercell already yields relatively well converged results.

\begin{table}
  \caption{\textbf{Convergence of vertical excitation energies (VEEs) with respect to the size of the supercell.} VEEs (eV) of many-body excited states relative to the $\prescript{3}{}{A}_2$ ground state of the NV$^-$ center in diamond computed using TDDFT at the PBE level of theory. Results obtained with the $(2 \times 2 \times 2)$, the $(3 \times 3 \times 3)$ and the $(4 \times 4 \times 4)$ supercells are shown.}
  \label{s-tab:size}
  \begin{ruledtabular}
  \begin{tabular}{cccc}
    Excited state & $(2\times2\times2)$ & $(3\times 3 \times 3)$ & $(4\times 4 \times 4)$ \\
    \hline
    $\prescript{3}{}{E}$ & 1.978 & 2.076 & 2.095 \\
    $\prescript{1}{}{A}_1$ & 1.283 & 1.334 & 1.341 \\
    $\prescript{1}{}{E}$ & 0.504 & 0.510 & 0.512 \\
  \end{tabular}
  \end{ruledtabular}
\end{table}

\section*{\label{s-sec:vee} Supplementary Note 2: Analysis of vertical excitation energies}

\begin{table}
  \caption{\textbf{Contribution of Slater determinants to the many-body electronic states of the NV$^-$ center in diamond.} Contribution of single and double excitation Slater determinants (\%) to the many-body electronic states, as obtained from  TDDFT and QDET calculations. Slater determinants are denoted in the hole notation, and the contributions are given in terms of the coefficients defined in equation~(1) of the main text.}
  \label{s-tab:wfxn}
  \begin{ruledtabular}
  \begin{tabular}{c|c|ccc|ccc|ccc}
    & & \multicolumn{3}{c|}{TDDFT@PBE} & \multicolumn{3}{c|}{TDDFT@DDH} & \multicolumn{3}{c}{QDET} \\ \hline 
    & Slater det. &$\prescript{1}{}{E}_x$ & $\prescript{1}{}{E}_y$ & $\prescript{1}{}{A}_1$ & $\prescript{1}{}{E}_x$ & $\prescript{1}{}{E}_y$ & $\prescript{1}{}{A}_1$ & $\prescript{1}{}{E}_x$ & $\prescript{1}{}{E}_y$ & $\prescript{1}{}{A}_1$ \\
    \hline
    & $|e_x\overline{e}_x\rangle$ & 45.1 & -- & 50.0 & 41.9 & -- & 49.8 & 44.9 & -- & 46.9 \\
    & $|e_x\overline{e}_y\rangle$ & -- & 45.1 & -- & -- & 41.9 & -- & -- & 44.9 & -- \\
    Single & $|e_y\overline{e}_x\rangle$ & -- & 45.1 & -- & -- & 41.9 & -- & -- & 44.9 & -- \\
    Excitations & $|e_y\overline{e}_y\rangle$ & 45.1 & -- & 50.0 & 41.9 &  -- & 49.8 & 44.9 & -- & 46.9 \\
    & $|a_1\overline{e}_x\rangle$ & 4.0 & 1.1 & -- & 5.9 & -- & -- & 3.3 & -- & -- \\
    & $|a_1\overline{e}_y\rangle$ & 1.1 & 4.0 & -- & -- & 5.9 & -- & -- & 3.3 & -- \\\hline
    Double & $|e_x\overline{a}_1\rangle$ & -- & -- & -- & -- & -- & -- & 3.3 & -- & -- \\
    Excitations & $|e_y\overline{a}_1\rangle$ & -- & -- & -- & -- & -- & -- & -- & 3.3 & -- \\
    & $|a_1\overline{a}_1\rangle$ & -- & -- & -- & -- & --& --& --& --& 2.6 \\
  \end{tabular}
  \end{ruledtabular}
\end{table}

As discussed in the main text, we attribute the overestimate of the VEE of the $\prescript{1}{}{A}_1$ state to the neglect of the double excitation Slater determinant $|a_1\overline{a}_1\rangle$. Here, we provide a simple estimate of the error based on perturbation theory: the coupling strength between the double excitation Slater determinant and the singlet state, $\langle a_1\overline{a}_1 | \hat{H} | \prescript{1}{}{A}_1 \rangle$, can be estimated using the coefficient of the configuration $|a_1\overline{a}_1\rangle$, $c_{a_1\overline{a}_1}^{\prescript{1}{}{A}_1}$ predicted by QDET calculations~\cite{nan2022dct}, and the energy difference, $E_{\prescript{1}{}{A}_1} - E_{a_1\overline{a}_1}$; the correction to the energy of the singlet state, $E^{(2)}_{\prescript{1}{}{A}_1}$, can then be estimated using the coupling strength and the difference $E_{\prescript{1}{}{A}_1} - E_{a_1\overline{a}_1}$. Here $\hat{H}$ is the many-body electronic Hamiltonian.

QDET results~\cite{nan2022dct} indicate that the weight of the $|e_x\overline{e}_x\rangle$ and $|e_y\overline{e}_y\rangle$ configurations in the $\prescript{1}{}{A}_1$ state is $\left|c_{e_x\overline{e}_x}^{\prescript{1}{}{A}_1}\right|^2 + \left|c_{e_y\overline{e}_y}^{\prescript{1}{}{A}_1}\right|^2 = 0.938$ (see Supplementary Table~\ref{s-tab:wfxn}), while the weight of $|a_1\overline{a}_1\rangle$ and other configurations with higher energies amounts to 0.062~\cite{nan2022dct}. For simplicity we assume the weight of $|a_1\overline{a}_1\rangle$ configuration is 0.062, corresponding to $\left|c_{a_1\overline{a}_1}^{\prescript{1}{}{A}_1}\right| = 0.25$~\cite{nan2022dct}. The energy of the $|a_1\overline{a}_1\rangle$ configuration relative to the $\prescript{3}{}{A}_2$ ground state is 7.33 eV, estimated using Kohn-Sham orbital energies at the DDH level of theory. It is thus 5.36 eV higher than the $\prescript{1}{}{A}_1$ state. Within perturbation theory, the coefficient of the $|a_1\overline{a}_1\rangle$ configuration is
\begin{equation}
    \left|c_{a_1\overline{a}_1}^{\prescript{1}{}{A}_1}\right| \approx 0.25 \approx \bigg|\frac{\langle a_1\overline{a}_1  | \hat{H} |\prescript{1}{}{A}_1 \rangle}{E_{\prescript{1}{}{A}_1} - E_{a_1\overline{a}_1}}\bigg|.
\end{equation}
With the estimated $E_{\prescript{1}{}{A}_1} - E_{a_1\overline{a}_1} = -5.36 $ eV, we have $|\langle a_1\overline{a}_1  | \hat{H} |\prescript{1}{}{A}_1 \rangle| \approx 1.34$ eV. Therefore the second order correction to the energy is
\begin{equation}
    E^{(2)}_{\prescript{1}{}{A}_1} = \frac{|\langle a_1\overline{a}_1  | \hat{H} |\prescript{1}{}{A}_1 \rangle|^2}{E_{\prescript{1}{}{A}_1} - E_{a_1\overline{a}_1}} \approx -0.34\ \text{eV}.
\end{equation}
Using this correction to estimate the energy of the $\prescript{1}{}{A}_1$ state we obtain 1.64 eV, which is in good agreement with the experimentally inferred zero-phonon absorption energy of $1.51 \sim 1.60 $ eV~\cite{rogers2008infrared} and the QDET VEE value of 1.270 eV~\cite{nan2022dct}.

Similarly, including the $|e_x\overline{a}_1\rangle$ Slater determinant in the expansion of the many-body wavefunction brings the VEE of the $\prescript{1}{}{E}_x$ state closer to the experimental and high-level theoretical results. For the $\prescript{1}{}{E}_x$ state, we have $E_{\prescript{1}{}{E}_x} - E_{e_x\overline{a}_1} = - 5.53$ eV, and $\left|c_{e_x\overline{a}_1}^{\prescript{1}{}{E}_x}\right| = 0.18$ from QDET results. Hence we can estimate $\langle e_x\overline{a}_1 | \hat{H} | \prescript{1}{}{E}_x \rangle \approx 1.00$ eV and an energy correction $E^{(2)}_{\prescript{1}{}{E}_x} \approx -0.18$ eV. Using this energy correction, the energy of the $\prescript{1}{}{E}_x$ state becomes 0.50 eV, in much better agreement with the experimentally inferred zero-phonon absorption energy of $0.34 \sim 0.43$ eV~\cite{goldman2015isc,goldman2017erratum} and the QDET VEE value of 0.463 eV~\cite{nan2022dct}. For the $\prescript{1}{}{E}_y$ state, including the $|e_y\overline{a}_1\rangle$ configuration brings the value of VEE to 0.50 eV.

For completeness we note that QDET and TDDFT calculations are built on spin-unpolarized and spin-polarized ground state Kohn-Sham orbitals, respectively, and this difference has been neglected in our comparison.

\section*{\label{s-sec:dis} Supplementary Note 3: Geometry Relaxation}

\setlength{\footnotesep}{0.6cm}
\begin{table*}
  \caption{\textbf{Geometry relaxation in excited states.} Mass-weighted displacement $\Delta Q$ (amu$^{0.5}$ \AA) between the equilibrium geometries of the $\prescript{3}{}{A}_2$ ground state (GS) and the excited states (ES), and the Franck-Condon (FC) shift $E_{\text{FC}}$ (meV) in the ESs. Equilibrium geometries of the ESs are obtained using TDDFT or constrained occupations DFT ($\Delta$SCF) with PBE and DDH functionals.}
  \label{s-tab:relax}
  \begin{ruledtabular}
  \begin{tabular}{ccccc}
  \centering
    State & Method & Configuration\footnote{Electronic configuration of many-body states in the hole notation.} & $\Delta Q$\footnote{$\Delta Q = \left(\sum_{\alpha=1}^{N_{\text{atoms}}} \sum_{i=x,y,z}M_\alpha (\mathbf{R}_{\alpha i, \text{ES}} - \mathbf{R}_{\alpha i, \text{GS}})^2 \right)^{0.5}$, where $M_{\alpha}$ is the mass of the $\alpha$th atom, $\mathbf{R}_{\alpha i, \text{ES}}$ ($\mathbf{R}_{\alpha i, \text{GS}}$) is the equilibrium atomic structure of the ES (GS).} & $E_{\text{FC}} \footnote{$E_{\text{FC}} = E_{\text{ES}@\text{GS}} - E_{\text{ES}@\text{ES}}$, where $E_{\text{ES}@\text{GS}}$ is the total energy of ES at the equilibrium geometry of the GS, and $E_{\text{ES}@\text{ES}}$ is the total energy of ES at the equilibrium geometry of the ES.}$ \\
    \hline
    $\prescript{3}{}{E}$ & TDDFT (PBE) & $\overline{a}_1\overline{e}$ & 0.583 & 187 \\
    $\prescript{3}{}{E}$ & TDDFT (DDH) & $\overline{a}_1\overline{e}$ & 0.634 & 255 \\
    $\prescript{3}{}{E}$ & $\Delta$SCF (PBE) & $\overline{a}_1\overline{e}$ & 0.625 & 219 \\
    $\prescript{3}{}{E}$ & $\Delta$SCF (DDH)\footnote{The $\overline{a}_1\overline{e}_x^{0.5}\overline{e}_y^{0.5}$ configuration was used to converge the $\Delta$SCF (DDH) calculation.} & $\overline{a}_1\overline{e}$ & 0.635 & 246 \\
    $\prescript{1}{}{A}_1$ & TDDFT (PBE) & $e\overline{e}$ & 0.111 & 17 \\
    $\prescript{1}{}{A}_1$ & TDDFT (DDH) & $e\overline{e}$ & 0.094 & 17 \\
    $\prescript{1}{}{E}$ & TDDFT (PBE) & $e\overline{e}$ & 0.415 & 64 \\
    $\prescript{1}{}{E}$ & TDDFT (DDH) & $e\overline{e}$ & 0.413 & 105 \\
  \end{tabular}
  \end{ruledtabular}
\end{table*}

To quantify the magnitude of geometry relaxation in excited states, we computed the mass-weighted atomic displacement between geometries of excited states and the $\prescript{3}{}{A}_2$ ground state, $\Delta Q$, and the Franck-Condon (FC) shift, $E_{\text{FC}}$, for many-body excited states, as shown in Supplementary Table~\ref{s-tab:relax}. TDDFT results obtained at the PBE level are generally similar with those obtained at the DDH level. For the $\prescript{3}{}{E}$ excited state, TDDFT and constrained-occupations DFT ($\Delta$SCF) yield comparable results. For the $\prescript{1}{}{A}_1$ singlet state, TDDFT predicts a minor displacement and FC shift, indicating that the geometry of the $\prescript{1}{}{A}_1$ state is close to that of the $\prescript{3}{}{A}_2$ ground state. On the other hand, a more significant displacement and FC shift are found for the $\prescript{1}{}{E}$ state, indicating that its geometry is significantly displaced from that of the $\prescript{3}{}{A}_2$ ground state.

\section*{\label{s-sec:pjt-djt} Supplementary Note 4: Non-adiabatic coupling between singlet states}

\begin{figure}[t]
\includegraphics[width=16cm]{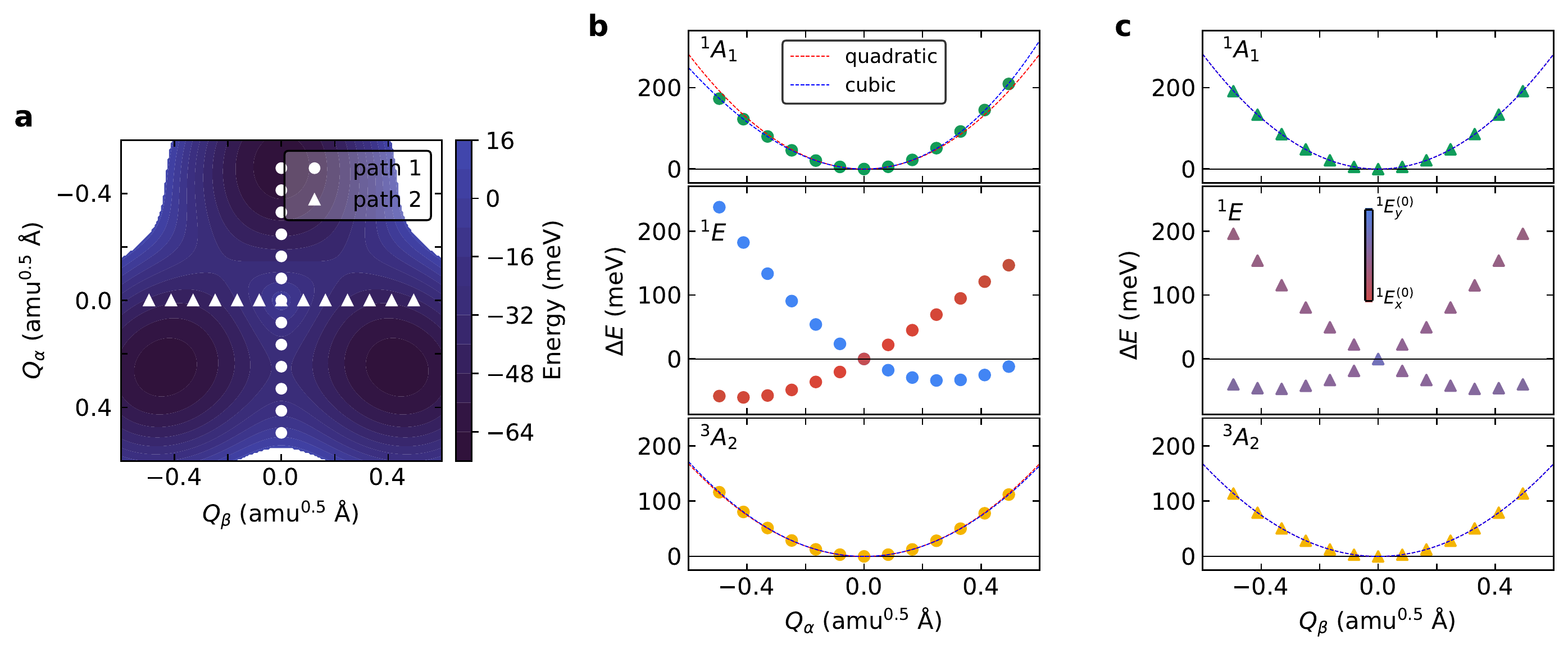}
\caption{\label{s-fig:pecs} \textbf{Potential energy curves (PECs) of the NV$^-$ center in diamond.} \textbf{a} Adiabatic potential energy surface (PES) of the lower branch of the $\prescript{1}{}{E}$ state. $Q_\alpha$, $Q_\beta$ configuration coordinates represent the collective motion of effective $e$ phonon modes. \textbf{b-c} PECs of the $\prescript{1}{}{A}_1$ state (up panel), the $\prescript{1}{}{E}$ states (middle panel) and the $\prescript{3}{}{A}_2$ state (bottom panel) along path 1 and path 2 defined in \textbf{a}. Relative change in total energy, $\Delta E$, computed at the PBE level of theory as a function of $Q_\alpha$ and $Q_\beta$ is reported. The color of the $\prescript{1}{}{E}$ states represents the weight of $\prescript{1}{}{E}_x^{(0)}$ and $\prescript{1}{}{E}_y^{(0)}$ configuration. The PECs of the $\prescript{1}{}{A}_1$ and $\prescript{3}{}{A}_2$ states are fitted by quadratic functions (red dashed lines) and cubic functions (blue dashed lines).}
\end{figure}

The electron-phonon coupling entering the effective Hamiltonian discussed in the main text reads
\begin{equation}
    \hat{H}_{e-ph} = \sum_{ij}\sum_{\lambda=\alpha,\beta} g_{ij,\lambda} \hat{c}_i^{\dagger} \hat{c}_j \left( \hat{b}_{\lambda}^{\dagger} + \hat{b}_{\lambda} \right),
\end{equation}
where $g_{ij,\lambda}$ is the linear electron-phonon coupling strength between electronic state $i$, $j$ and phonon mode $\lambda$. Considering $C_{3v}$ symmetry and using $e_x$ and $e_y$ vibrational modes as collective variables (CVs, which we call $Q_\alpha$, $Q_\beta$), the electron-phonon coupling strength can be simplified as
\begin{equation}
    g_{ij,\alpha} = \left(
    \begin{matrix}
    0 & \widetilde{G} & 0 \\
    \widetilde{G} & G & 0 \\
    0 & 0 & -G \\
    \end{matrix}
    \right), \quad g_{ij,\beta} = \left(
    \begin{matrix}
    0 & 0 & \widetilde{G} \\
    0 & 0 & -G \\
    \widetilde{G} & -G & 0 \\
    \end{matrix}
    \right).
\end{equation}
$\widetilde{G}$ is the coupling strength between the $\prescript{1}{}{A}_1$ and the $\prescript{1}{}{E}$ states,
\begin{equation}
    \widetilde{G} = \sqrt{\frac{\hbar}{2\omega_e}} \bigg\langle \prescript{1}{}{A}_1 \bigg | \frac{\partial V}{\partial Q_{\alpha}} \bigg | \prescript{1}{}{E}_{x} \bigg \rangle = \sqrt{\frac{\hbar}{2\omega_e}} \bigg\langle \prescript{1}{}{A}_1 \bigg | \frac{\partial V}{\partial Q_{\beta}} \bigg | \prescript{1}{}{E}_{y} \bigg \rangle,
\end{equation}
and $G$ is the coupling strength between the $\prescript{1}{}{E}_x$ and $\prescript{1}{}{E}_y$ states,
\begin{equation}
    G = \sqrt{\frac{\hbar}{2\omega_e}}\bigg \langle \prescript{1}{}{E}_x \bigg | \frac{\partial V}{\partial Q_{\alpha}} \bigg | \prescript{1}{}{E}_x \bigg \rangle = - \sqrt{\frac{\hbar}{2\omega_e}}\bigg \langle \prescript{1}{}{E}_y \bigg | \frac{\partial V}{\partial Q_{\alpha}} \bigg | \prescript{1}{}{E}_y \bigg \rangle = - \sqrt{\frac{\hbar}{2\omega_e}}\bigg \langle \prescript{1}{}{E}_x \bigg | \frac{\partial V}{\partial Q_{\beta}} \bigg | \prescript{1}{}{E}_y \bigg \rangle.
\end{equation}
Here $V$ is the potential energy of the many-body system. The dependence of $\widetilde{G}$ and $G$ on $Q_\alpha$ and $Q_\beta$ is neglected.

We computed the potential energy curves (PECs) of many-body electronic states along path 1 ($Q_\beta = 0$) and path 2 ($Q_\alpha = 0$) using TDDFT with the PBE functional, as shown in Supplementary Fig.~\ref{s-fig:pecs}. Along path 1, we observe a local minimum for $Q_\alpha < 0$, a cusp at $Q_\alpha = 0$ and 60 meV higher than the minimum, and a saddle point for $Q_\alpha > 0$ with a barrier height of 26 meV. The PEC of the $\prescript{1}{}{A}_1$ state has a curvature that corresponds to an effective phonon energy of 81 meV, which is 19 meV higher than that of the $\prescript{3}{}{A}_2$ ground state. The PEC of the $\prescript{1}{}{A}_1$ state is also slightly anharmonic as shown by the quality of a cubic function fit (blue dashed line) compared to a quadratic one (red dashed line).

By fitting the adiabatic PECs along path 1 and path 2 obtained by solving the effective Hamiltonian, to the PECs from first-principles calculations, we obtained the parameters for the effective Hamiltonian, i.e. $\hbar\omega_e = 63.0$ meV, $\widetilde{G} = 133.2$ meV, and $G = 62.4$ meV, as shown in Supplementary Fig.~\ref{s-fig:fit-PEC}.

\begin{figure*}
    \centering
    \includegraphics[width=16cm]{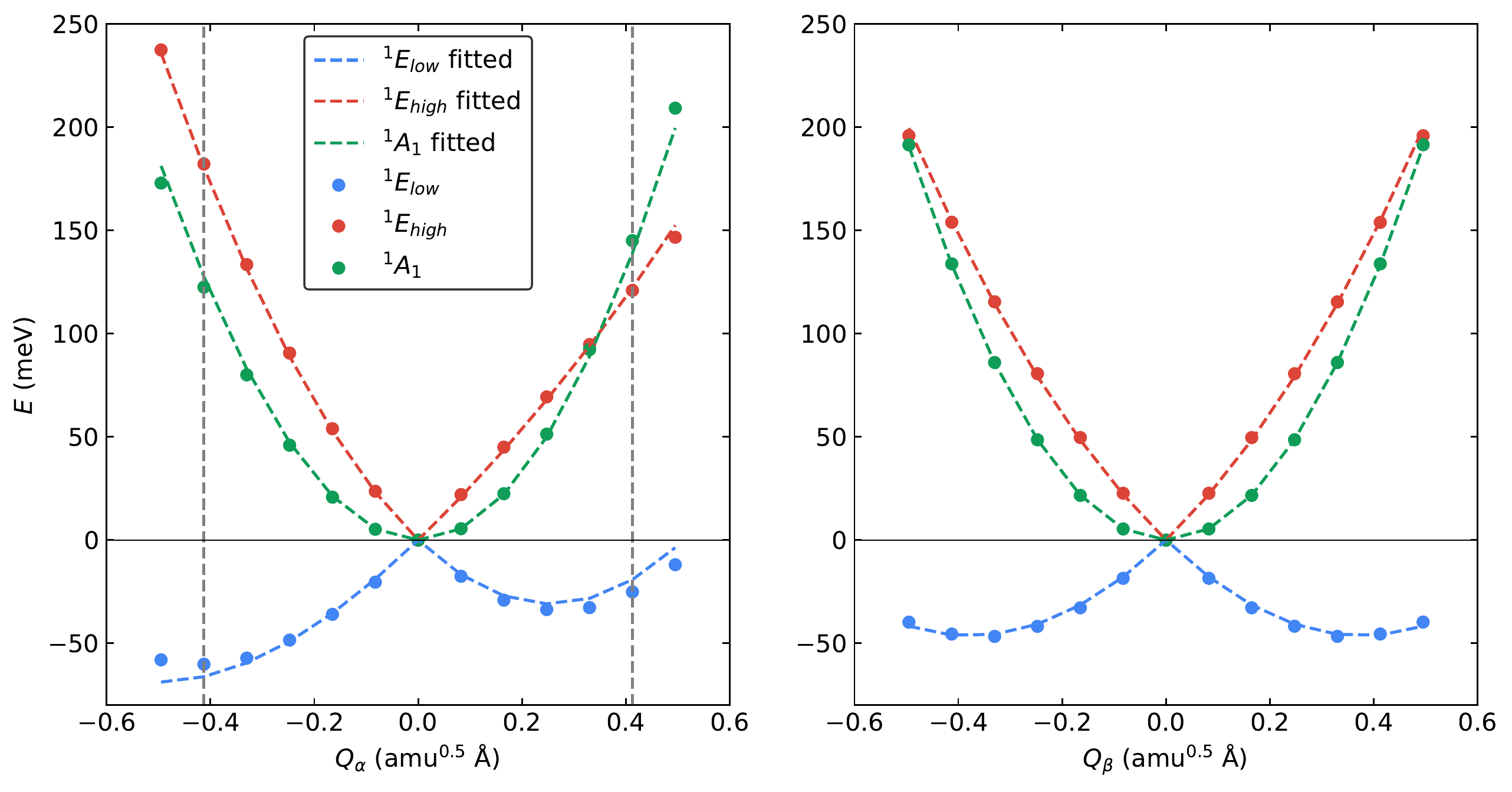}
    \caption{\textbf{Fitting of the potential energy curves (PECs) of the NV$^-$ center in diamond.} Potential energy curves (PECs) are obtained from first-principles calculations (dots) and by solving the effective Hamiltonian (dashed lines) along path 1 and path 2, as defined in Supplementary Fig.~\ref{s-fig:pecs}. Minima of the $\prescript{1}{}{A}_1$ state are arbitrarily shifted to 0 meV. The parameters of the effective Hamiltonian (see equation (2) of the main text) are obtained by best fitting the first-principles results.}
    \label{s-fig:fit-PEC}
\end{figure*}
\begin{figure*}
    \centering
    \includegraphics[width=16cm]{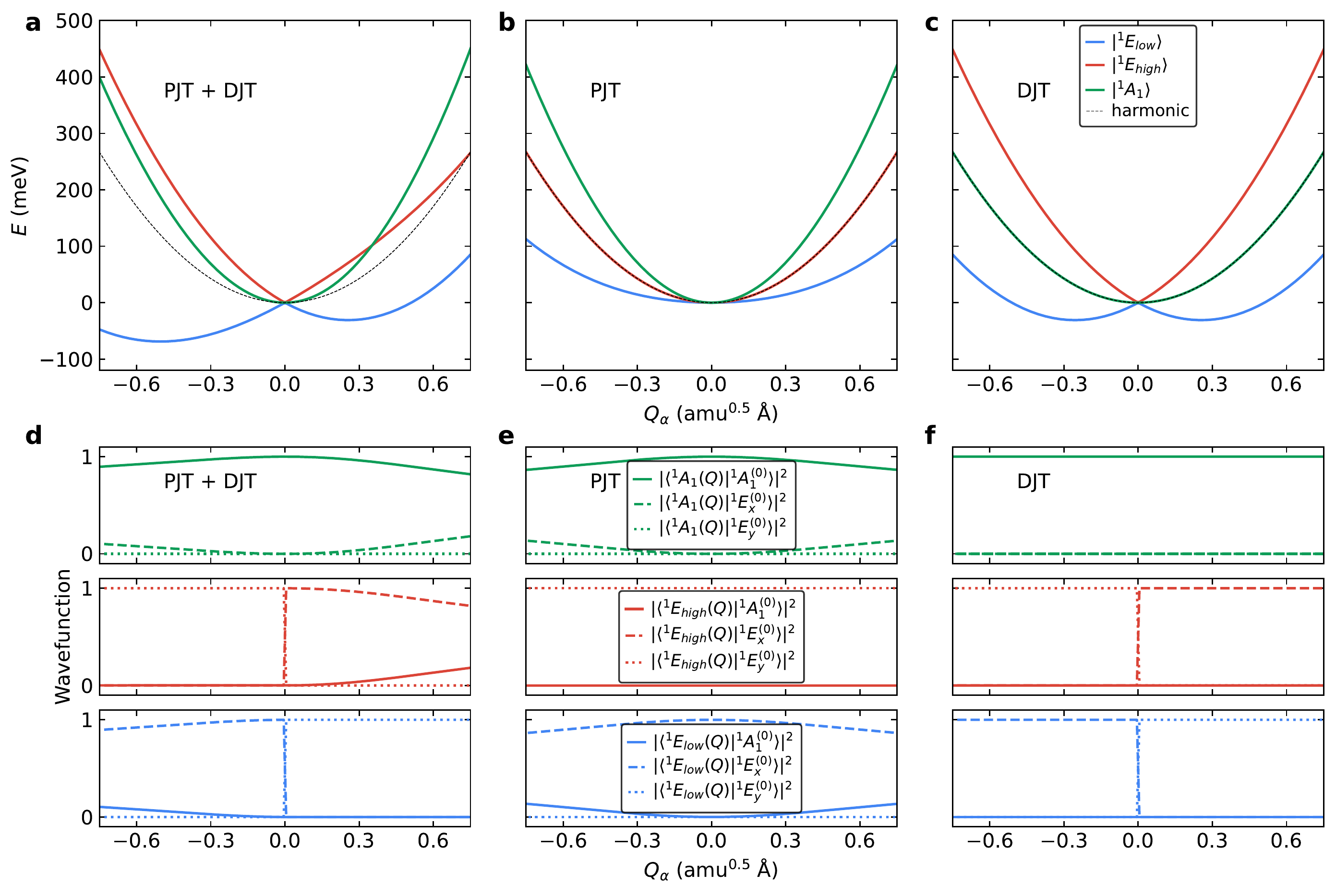}
    \caption{\textbf{Dynamic Jahn-Teller (DJT) and pseudo Jahn-Teller (PJT) effect.} Potential energy curves (PECs) of the $\prescript{1}{}{A}_1$ and $\prescript{1}{}{E}$ states along the path with $Q_\beta = 0$ obtained by solving the effective Hamiltonian for non-adiabatic coupling, considering (\textbf{a}) both the PJT and the DJT effect, (\textbf{b}) only the PJT effect, and (\textbf{c}) only the DJT effect. Minima of the $\prescript{1}{}{A}_1$ curve are arbitrarily shifted to 0 meV for a better comparison. Decomposition of the wavefunctions of the $\prescript{1}{}{A}$ and $\prescript{1}{}{E}$ states onto the wavefunctions at $Q_\alpha = 0$ are displayed in the bottom panel, considering (\textbf{d}) both the PJT and the DJT effect, (\textbf{e}) only the PJT effect, and (\textbf{f}) only the DJT effect.}
    \label{s-fig:pjt-djt-1}
\end{figure*}
\begin{figure*}
    \centering
    \includegraphics[width=16cm]{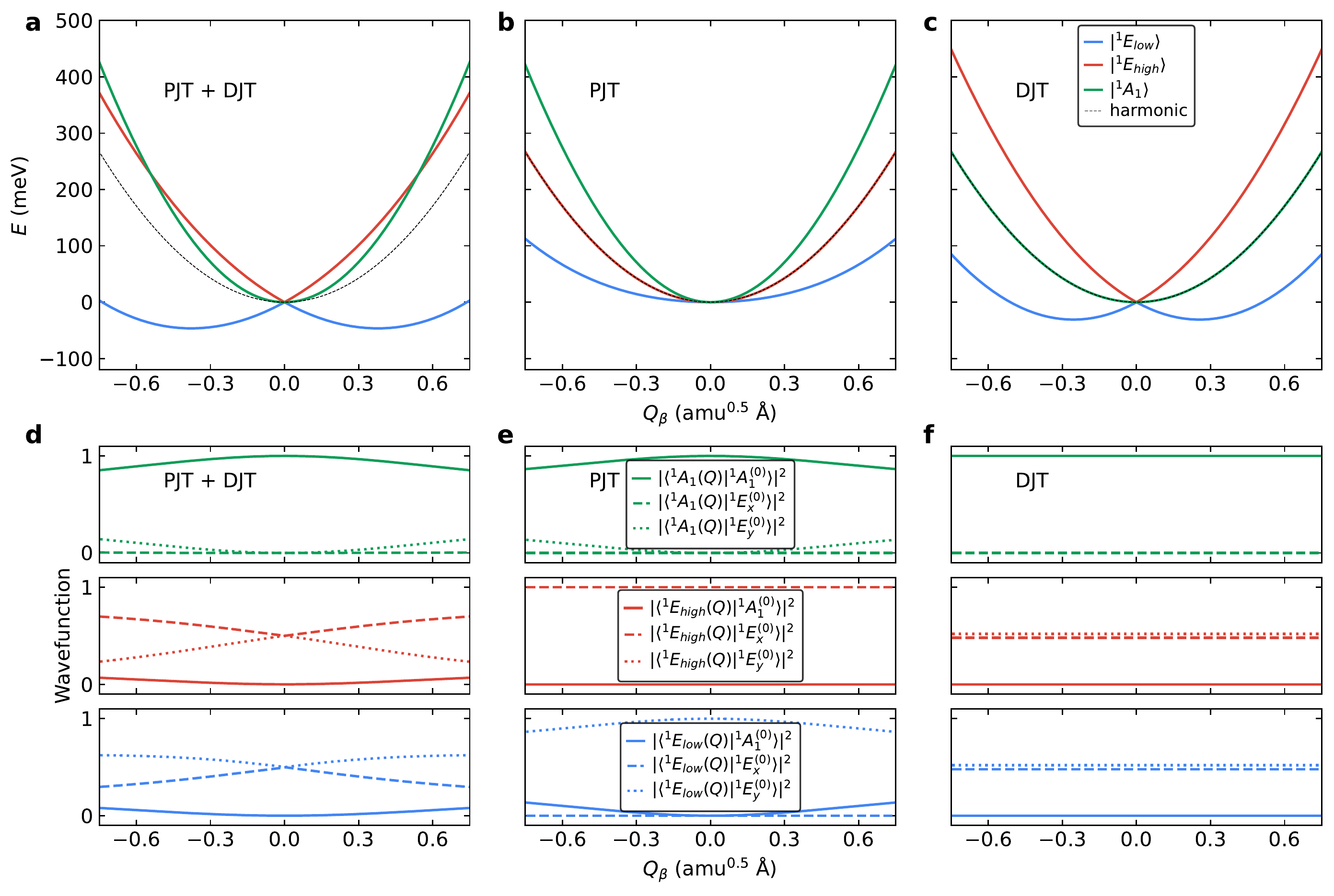}
    \caption{\textbf{Dynamic Jahn-Teller (DJT) and pseudo Jahn-Teller (PJT) effect.} Potential energy curves (PECs) of the $\prescript{1}{}{A}_1$ and $\prescript{1}{}{E}$ states along the path with $Q_\alpha = 0$ considering (\textbf{a}) both the PJT and the DJT effect, (\textbf{b}) only the PJT effect, and (\textbf{c}) only the DJT effect. Minima of the $\prescript{1}{}{A}_1$ curve are arbitrarily shifted to 0 meV for a better comparison. Decomposition of the wavefunctions of the $\prescript{1}{}{A}_1$ and $\prescript{1}{}{E}$ states onto the wavefunctions at $Q_\beta = 0$ are displayed in the bottom panel, considering (\textbf{d}) both the PJT and the DJT effect, (\textbf{e}) only the PJT effect, and (\textbf{f}) only the DJT effect.}
    \label{s-fig:pjt-djt-2}
\end{figure*}

The non-adiabatic coupling can be analyzed in terms of the pseudo Jahn-Teller (PJT) interaction between the $\prescript{1}{}{A}_1$ and the $\prescript{1}{}{E}$ states with strength $\widetilde{G}$, together with the dynamic Jahn-Teller (DJT) interaction between two $\prescript{1}{}{E}$ states with strength $G$~\cite{bersuker2006jt, thiering2018nvisc}. Here we investigate the contribution of the PJT and DJT effect separately by setting either $G$ or $\widetilde{G}$ to zero along path 1 (Supplementary Fig.~\ref{s-fig:pjt-djt-1}) and path 2 (Supplementary Fig.~\ref{s-fig:pjt-djt-2}), respectively. The DJT effect only couples two $\prescript{1}{}{E}$ states. Along path 1 (path 2), the DJT effect shifts the local minima of the $\prescript{1}{}{E}$ states away from the $Q_\alpha = 0$ ($Q_\beta = 0$) point and results in the presence of a cusp. On the other hand, the PJT effect couples $\prescript{1}{}{A}_1$ and $\prescript{1}{}{E}$ states. It increases the curvature of the $\prescript{1}{}{A}_1$ PES; it also decreases the curvature of the lower branch of the $\prescript{1}{}{E}$ PESs that can be characterized as $\prescript{1}{}{E}_x$ ($\prescript{1}{}{E}_y$) state along path 1 (path 2), and is accompanied by the mixing of $\prescript{1}{}{E}_x$ ($\prescript{1}{}{E}_y$) wavefunctions with the $\prescript{1}{}{A}_1$ wavefunction. Either the PJT or the DJT effect taken separately can preserve the axial symmetry of the PESs; Preservation of the axial symmetry is the typical result when electron-phonon coupling is considered only at the linear order. When the PJT and the DJT effects are considered at the same time, the axial symmetry of the PESs is reduced to $C_{3v}$ symmetry.

\section*{\label{s-sec:hrf&dwf} Supplementary Note 5: Huang-Rhys factors and spectral densities}
\begin{table}
  \caption{\textbf{Computed Huang-Rhys factors (HRFs) and Debye-Waller factors (DWFs).} Total HRFs and DWFs (\%) for $\prescript{3}{}{E} \to \prescript{3}{}{A}_2$ and $\prescript{1}{}{E} \to \prescript{1}{}{A_1}$ transitions computed at different levels of theory are compared with experiments~\cite{kehayias2013nvexp}. Here DDH$-\Delta Q$, PBE$-ph$ denotes results based on the Huang-Rhys theory, evaluated using $\Delta Q$ computed at the DDH level and the phonon modes computed at the PBE level of theory. The DWF is calculated as $\text{DWF = }e^{-\text{HRF}}$. These results are obtained in the dilute limit, approximated by a $(12 \times 12 \times 12)$ supercell with 13824 atomic sites.}
  \label{s-tab:hrfdwf}
  \begin{ruledtabular}
  \begin{tabular}{ccccc}
    & \multicolumn{2}{c}{$\prescript{3}{}{E} \to \prescript{3}{}{A}_2$} & \multicolumn{2}{c}{$\prescript{1}{}{E} \to \prescript{1}{}{A_1}$}  \\
    & HRF & DWF(\%) & HRF & DWF(\%) \\
    \hline
    PBE$-\Delta Q$, PBE$-ph$ & 2.44 & 8.7 & 1.42 & 24 \\
    DDH$-\Delta Q$, PBE$-ph$ & 2.97 & 5.1 & 1.08 & 34 \\
    DDH$-\Delta Q$, DDH$-ph$ & 3.08 & 4.6 & 1.13 & 32 \\
    Expt.~\cite{kehayias2013nvexp} & 3.49 & 3.2 & $\sim$0.9 & $\sim$40 \\
  \end{tabular}
  \end{ruledtabular}
\end{table}

We compared the Huang-Rhys factors (HRFs) and Debye-Waller factors (DWFs) obtained using TDDFT at the PBE and the DDH level, as shown in Supplementary Table~\ref{s-tab:hrfdwf}. Due to the unaffordable cost of computing phonon modes at the DDH level of theory, we approximated the DDH phonons as PBE phonons whose phonon energies are multiplied by a constant factor~\cite{jin2021pl}. For both the $\prescript{3}{}{E} \to \prescript{3}{}{A}_2$ transition and the $\prescript{1}{}{E} \to \prescript{1}{}{A}_1$ transition, the best agreement with experiment is obtained when the displacement computed at the DDH level of theory (DDH$-\Delta Q$) was used. Using phonons computed at the PBE level of theory (PBE$-ph$) or DDH level of theory (DDH$-ph$) only results in minor changes.

\begin{figure*}
\includegraphics[width=16 cm]{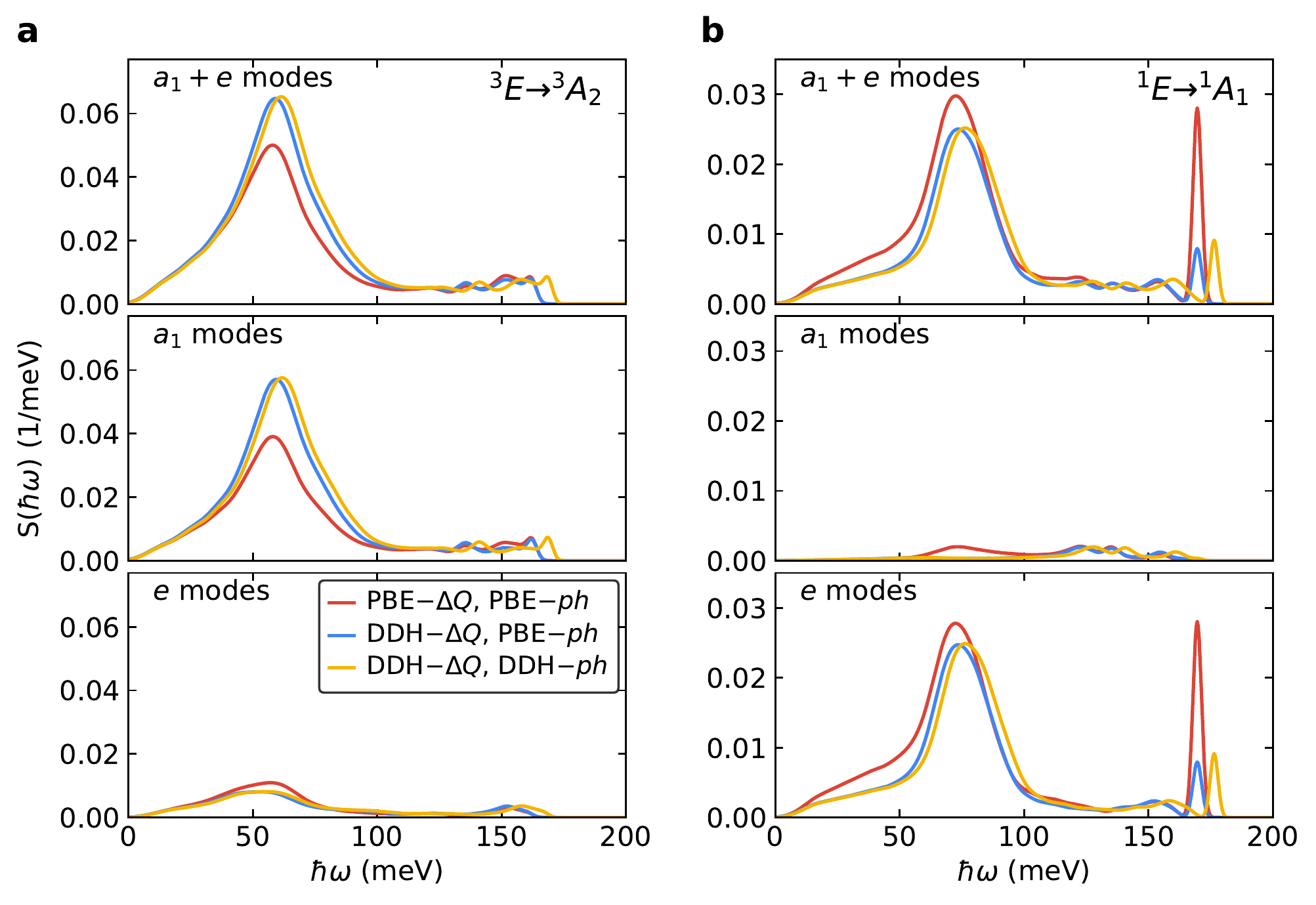}
\caption{\label{s-fig:comp-hr}\textbf{Spectral densities of electron-phonon coupling.} Spectral densities of electron-phonon coupling $S(\hbar\omega)$ for the $\prescript{3}{}{E} \to \prescript{3}{}{A}_2$ transition (\textbf{a}) and the $\prescript{1}{}{E} \to \prescript{1}{}{A}_1$ transition (\textbf{b}). Contributions of the $a_1$ type and $e$ type phonon modes are shown in the middle and the bottom panels, respectively. Here DDH$-\Delta Q$, PBE$-ph$ denotes results based on the Huang-Rhys theory, evaluated using $\Delta Q$ computed at the DDH level and the phonon modes computed at the PBE level of theory. These results are obtained in the dilute limit, approximated by a $(12 \times 12 \times 12)$ supercell with 13824 atomic sites.}
\end{figure*}

Spectral density of electron-phonon coupling, $S(\hbar\omega)$, computed at different levels of theory are shown in Supplementary Fig.~\ref{s-fig:comp-hr}. Peak positions of the spectral densities computed with DDH$-ph$ are generally right shifted relative to those obtained with PBE. For the $\prescript{3}{}{E} \to \prescript{3}{}{A}_2$ transition, all spectral densities have a similar shape, while the one computed with DDH$-\Delta Q$ has higher intensities than the one computed with PBE$-\Delta Q$. The opposite is found for the $\prescript{1}{}{E} \to \prescript{1}{}{A}_1$ transition, where the spectral densities computed with DDH$-\Delta Q$ have smaller intensities than the one computed with PBE$-\Delta Q$. The spectral density for the $\prescript{1}{}{E} \to \prescript{1}{}{A_1}$ transition is dominated by the coupling with $e$ type phonons, while the one of the $\prescript{3}{}{E} \to \prescript{3}{}{A}_2$ transition is dominated by the coupling with the $a_1$ type phonons. When using PBE phonons, we find that the spectral density for the $\prescript{1}{}{E} \to \prescript{1}{}{A_1}$ transition has a broad (sharp) peak at 73 meV (170 meV), resulting from the coupling with a quasi-local (local) $e$ type phonon mode. Instead, the spectral density for the $\prescript{3}{}{E} \to \prescript{3}{}{A}_2$ transition mainly couples with the quasi-local $a_1$ type phonon mode, resulting in a broad peak at 60 meV. The 162 meV $a_1$ type local phonon mode is present in both the $\prescript{1}{}{A}_1$ and $\prescript{3}{}{A}_2$ states, but it does not contribute to the spectral density for the $\prescript{1}{}{E} \to \prescript{1}{}{A_1}$ transition. However, the 170 meV $e$ type local phonon mode is only present in the $\prescript{1}{}{A}_1$ state. The 170 meV $e$ mode has an energy higher than optical phonons of diamond and couples weakly to the vibrations of the diamond lattice. The spectral density of the $\prescript{1}{}{E} \to \prescript{1}{}{A_1}$ transition generally shifts to higher energy direction compared to that of  the $\prescript{3}{}{E} \to \prescript{3}{}{A_2}$ transition, implying an increase of phonon energies in the $\prescript{1}{}{A}_1$ state compared with those of the $\prescript{3}{}{A}_2$ state.

\begin{figure*}
\includegraphics[width=16 cm]{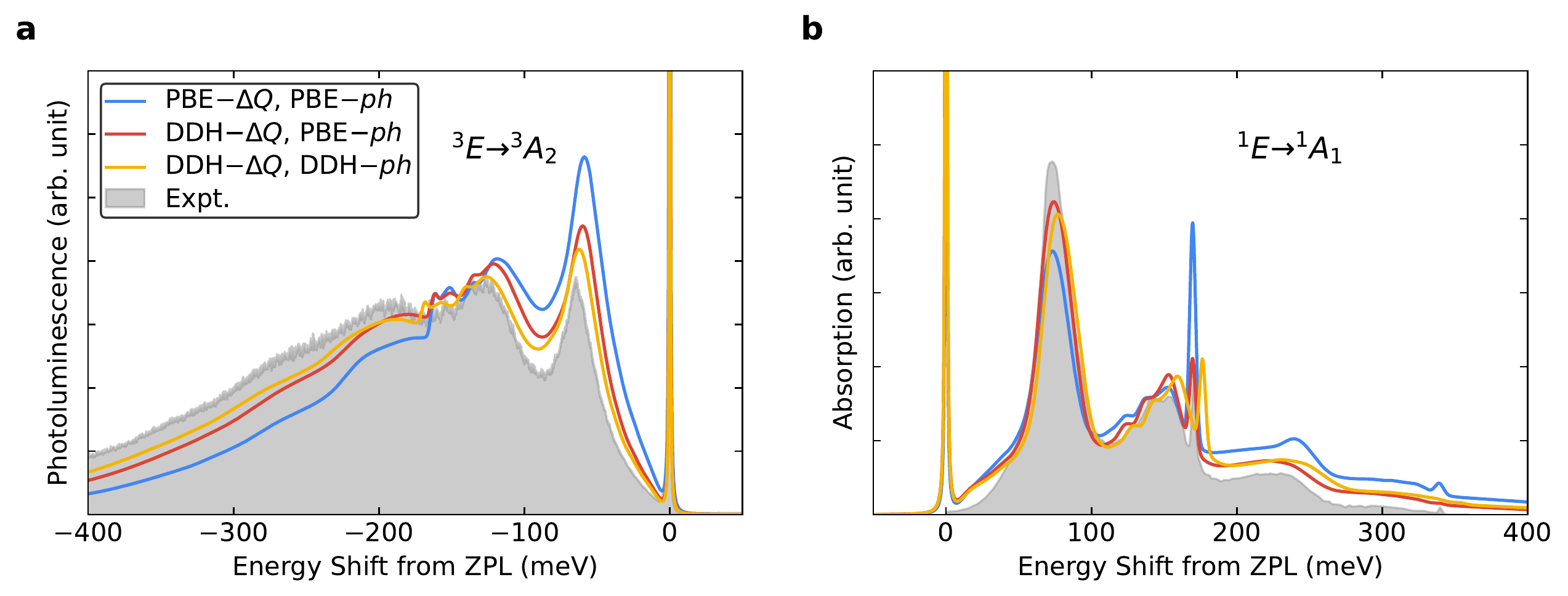}
\caption{\label{s-fig:comp-pl}\textbf{Optical spectra.} Photoluminescence (PL) spectra of the $\prescript{3}{}{E} \to \prescript{3}{}{A}_2$ transition (\textbf{a}) and absorption spectra of the $\prescript{1}{}{E} \to \prescript{1}{}{A}_1$ transition (\textbf{b}). Theoretical results are shown as solid lines while experimental results~\cite{kehayias2013nvexp,alkauskas2014luminescence} are shown as shaded areas. Here DDH$-\Delta Q$, PBE$-ph$ denotes the line shapes based on the Huang-Rhys theory, evaluated using $\Delta Q$ computed at the DDH level and the phonon modes computed at the PBE level of theory. These results are obtained in the dilute limit, approximated by a $(12 \times 12 \times 12)$ supercell with 13824 atomic sites.}
\end{figure*}

The photoluminescence (PL) spectrum of the $\prescript{3}{}{E} \to \prescript{3}{}{A}_2$ transition and the absorption spectrum of the $\prescript{1}{}{E} \to \prescript{1}{}{A}_1$ transition computed at difference levels of theory are shown in Supplementary Fig.~\ref{s-fig:comp-pl}. For both computed spectra, we obtained a better agreement with experiments when DDH$-\Delta Q$ was used. For the PL of the $\prescript{3}{}{E} \to \prescript{3}{}{A}_2$ transition, we obtained the best agreement with experiments when DDH$-ph$ was used, while for the absorption of the $\prescript{1}{}{E} \to \prescript{1}{}{A_1}$ transition, the best agreement was obtained when PBE$-ph$ was used. However, the difference caused by the choice of level of theory used for phonons and overall negligible.

\section*{\label{s-sec:phonon} Supplementary Note 6: Phonons of the $^3A_2$ and $\prescript{1}{}{A}_1$ states}

\begin{figure*}
\includegraphics[width=16 cm]{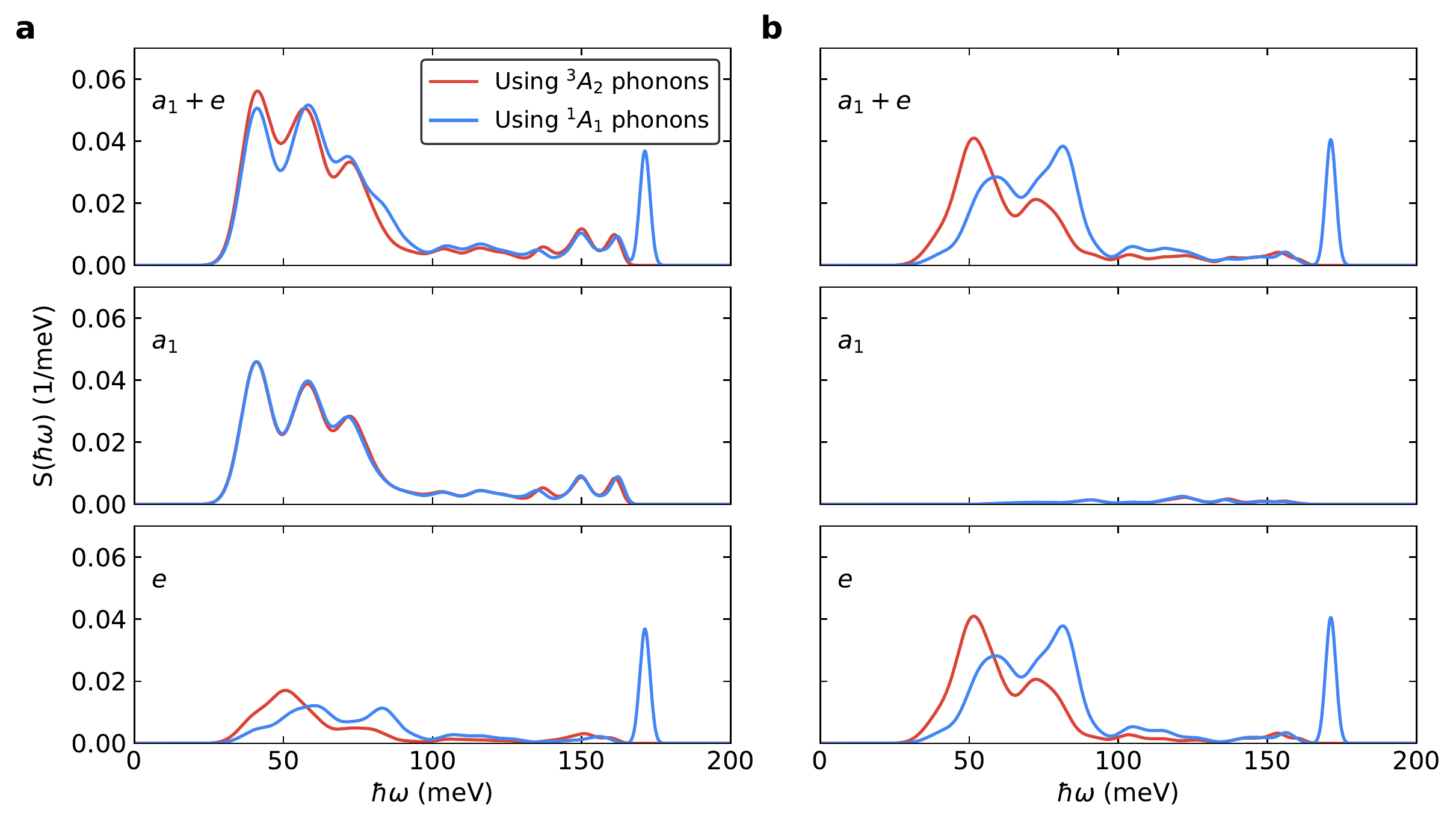}
\caption{\label{s-fig:el-ph} \textbf{Spectral densities of electron-phonon coupling computed using different sets of phonon modes.} Spectral densities $S(\hbar\omega)$ of the $\prescript{3}{}{E} \to \prescript{3}{}{A}_2$ transition (\textbf{a}) and the $\prescript{1}{}{E} \to \prescript{1}{}{A}_1$ transition (\textbf{b}). Red and blue lines are the spectral densities computed using the phonon modes of the $\prescript{3}{}{A}_2$ state and the $\prescript{1}{}{A}_1$ state, respectively. The entire spectral densities, contributions of the $a_1$ type phonons, and contributions of the $e$ type phonons are shown in the upper, middle and the bottom panels, respectively. The partial Huang-Rhys factors used to evaluate the spectral densities are computed using atomic displacements between different states, which are obtained at the PBE level of theory with the $(3\times3\times3)$ supercell with 216 atomic sites.}
\end{figure*}

To quantify the differences between the phonons of the $\prescript{3}{}{A}_2$ ground state and the phonons of the $\prescript{1}{}{A}_1$ singlet state, we computed the atomic displacements between equilibrium geometries obtained for the $\prescript{3}{}{E} \to \prescript{3}{}{A}_2$ and for the  $\prescript{1}{}{E} \to \prescript{1}{}{A}_1$ transitions. We then project the displacements onto the phonon modes of the $\prescript{3}{}{A}_2$ state and the $\prescript{1}{}{A}_1$ state and computed spectral densities of electron-phonon coupling $S(\hbar\omega)$, as shown in Supplementary Fig.~\ref{s-fig:el-ph}. For both the $\prescript{3}{}{E} \to \prescript{3}{}{A}_2$ and the $\prescript{1}{}{E} \to \prescript{1}{}{A}_1$ transition, the contributions of the $e$ type phonon modes shift to higher energies when the phonon modes of the $\prescript{1}{}{A}_1$ state are used, while the contributions of the $a_1$ type phonon modes remain almost unchanged. This finding implies that the $a_1$ type mode of the $\prescript{1}{}{A}_1$ and the $\prescript{3}{}{A}_2$ state are quite similar, which is consistent with the fact that the two states have similar geometries. The $e$-type mode of the $\prescript{1}{}{A}_1$ state has energies apparently higher than those of the $\prescript{3}{}{A}_2$ state, consistent with the analysis based on non-adiabatic coupling between singlet states.

\section*{\label{s-sec:anharmonicity} Supplementary Note 7: Treatment of anharmonic effects}

\begin{table}
  \caption{\textbf{Comparison of Huang-Rhys factors (HRFs) computed using forces and actual displacements.} Total HRFs of the $\prescript{1}{}{E} \to \prescript{1}{}{A}_1$ transition computed using either $\Delta Q_{k,F}$ ($\sum_k S_{k,F}$) or $\Delta Q_{k}$ ($\sum_k S_k$) for the $(3 \times 3\times 3)$ supercell. The ratio $r = \frac{a_{3}\Delta Q}{a_{2}}$ reported here is computed based on fitting the results of the one dimensional configurational coordinate diagrams (see Supplementary Fig.~\ref{s-fig:anharmonic}). Here DDH$-\Delta Q$, PBE$-ph$ denotes the line shapes based on the Huang-Rhys theory, evaluated using $\Delta Q$ computed at the DDH level and the phonon modes computed at the PBE level of theory.}
  \label{s-tab:hrf}
  \begin{ruledtabular}
  \begin{tabular}{ccc}
     &PBE$-\Delta Q$, PBE$-ph$ & DDH$-\Delta Q$, PBE$-ph$ \\
    \hline
    $\sum_k S_{k,F}$ & 1.33 & 1.03 \\
    $\sum_k S_k$ & 1.63 & 1.59 \\
    $\frac{\sum_k S_{k,F}}{\sum_k S_k}$ & 0.816 & 0.648 \\
    $r$ & -0.065 & -0.120 \\
    (1 + 3$r$) & 0.805 & 0.640 \\
  \end{tabular}
  \end{ruledtabular}
\end{table}

\begin{figure}
    \centering
    \includegraphics[width=10cm]{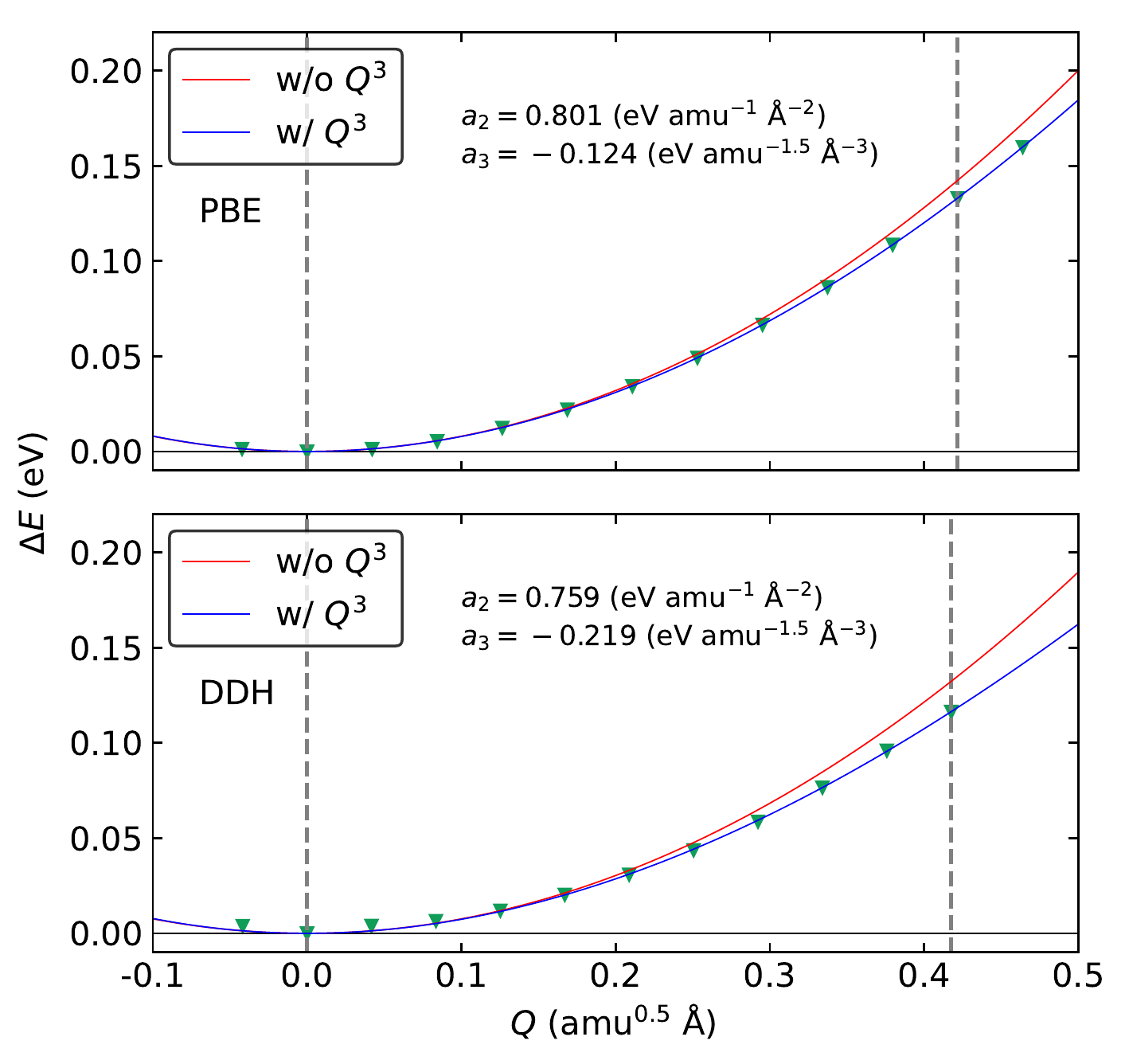}
    \caption{\textbf{Anharmonic effects of the potential energy curves (PECs).} The PECs of the $\prescript{1}{}{A}_1$ state along the linear path connecting the equilibrium geometries of the $\prescript{1}{}{A}_1$ state and the $\prescript{1}{}{E}$ state computed using TDDFT with PBE (top panel) and DDH (bottom panel) functional. The energy difference relative to the local minimum, $\Delta E$, is fitted as a function of $Q$ as $\Delta E = a_2Q^2 + a_3Q^3$. Blue lines are obtained with the fitted function, while the red lines are obtained neglecting the $a_3Q^3$ term.}
    \label{s-fig:anharmonic}
\end{figure}

As discussed in the main text, the non-adiabatic interaction between singlet states results in a noticeable anharmonicity of the PES of the $\prescript{1}{}{A}_1$ state, which needs to be carefully taken into account in the calculations of the absorption line shape of the $\prescript{1}{}{E} \to \prescript{1}{}{A}_1$ transition. The formalism used to describe the optical process relies on the HR theory, which assumes that the harmonic approximation holds. Since the anharmonic effect is small, we can still use the HR theory and introduce anharmonicity perturbatively. Following Ref.~\cite{zhu2009anh}, the spectral function for optical absorption including the lowest order of anharmonic effects reads:
\begin{equation}
    A_{\text{abs, an}}(\hbar\omega, T) = \frac{1}{2\pi\hbar} \int_{-\infty}^{\infty} e^{i\omega t} G_{\text{abs, an}} (t, T) e^{-\frac{\lambda |t|}{\hbar}} \mathrm{d} t,
\end{equation}
where $G_{\text{abs, an}} (\hbar\omega, T)$ is the generating function including the lowest order of anharmonic effects
\begin{equation}
\begin{aligned}
    G_{\text{abs, an}} (t, T) &= \exp \left[i t\left(2 \sum_{k} S_k \omega_k r_k \right) \right] \exp \bigg[\left(1 + 3 r_k \right) \bigg( \int_{-\infty}^{\infty} S(\hbar\omega) e^{-i\omega t} \mathrm{d} (\hbar\omega) - \sum_k S_k \\
    &+  \int_{-\infty}^{\infty} C(\hbar\omega, T) e^{-i\omega t} \mathrm{d} (\hbar\omega) + \int_{-\infty}^{\infty} C(\hbar\omega, T) e^{i\omega t} \mathrm{d} (\hbar\omega) - 2\sum_k \overline{n}_k(T)S_k \bigg) \bigg].
\end{aligned}
\end{equation}
Here we have expanded the potential energy to third order with respect to the $k$-th phonon mode 
\begin{equation}
\label{s-eq:potential}
    V_k(Q_k) = a_{k2} Q_k^2 + a_{k3} Q_k^3.
\end{equation}
and we have defined $r_k$ as the unitless ratio between the third and the second order coefficients of the expansion 
\begin{equation}
    r_k = \frac{a_{k3}}{a_{k2}} \Delta Q_k \,.
    \label{eq:r}
\end{equation}

It is evident that the anharmonicity affects the line shape in two ways. Firstly, it shifts the position of the zero-phonon line and also the phonon side bands from $E_{\text{ZPL}}$ to $E_{\text{ZPL}} - 2\sum_k S_k (\hbar\omega_k) r_k$. Further, it scales the HRFs. The anharmonic HR factor can be defined as
\begin{equation}
\label{s-eq:an-hrf}
    S_{k,\text{an}} = (1 + 3r_k) S_k.
\end{equation}
The latter effect is responsible for changing the line shape of the transition. 

In the following we show an efficient method to compute the anharmonic HR factor. We first compute, at the equilibrium atomic geometry of the $\prescript{1}{}{E}$ state, the force $\mathbf{F}_{\alpha}$ associated to the $\prescript{1}{}{A}_1$ PES and exerted on atom $\alpha$. We then compute the effective displacement associated to the $k$-th phonon,  $\Delta Q_{k,F}$, as
\begin{equation}
\begin{aligned}
    \Delta Q_{k,F} &= \frac{1}{\omega_k^2} \sum_{\alpha=1}^N  \frac{\mathbf{F}_{\alpha }\cdot \mathbf{e}_{k,\alpha}}{\sqrt{M_{\alpha}}}  = \frac{1}{\omega_k^2} \langle \widetilde{\mathbf{F}} | \mathbf{e}_k \rangle = \frac{\widetilde{\mathbf{F}}_k}{\omega_k^2},
    \label{eq:QkF}
\end{aligned}
\end{equation}
where $M_\alpha$ is the mass of the $\alpha$-th atom, and $\omega_k$  and $\mathbf{e}_{k,\alpha}$ are the energy and eigenvector of the $k$-th phonon mode, respectively. $\widetilde{\mathbf{F}}_\alpha = \mathbf{F}_\alpha / \sqrt{M_\alpha}$. If we use the definition of $\widetilde{\mathbf{F}}_k$ in terms of a derivative of the potential, and we express the potential as in equations~(\ref{s-eq:potential}), we obtain: 
\begin{equation}
    \widetilde{\mathbf{F}}_k = - \dfrac{\partial V_k(Q_k)}{\partial Q_k}\bigg|_{Q_k = \Delta Q_k} = - 2 a_{k2} \Delta Q_k - 3 a_{k3} \Delta Q_k^2 = - \omega_k^2 \Delta Q_k \left( 1 + \frac{3a_{k3}}{2a_{k2}} \Delta Q_k \right),
\end{equation}
where we used the fact that $a_{k2} = \frac{1}{2}\omega_k^2$. Here, $\Delta Q_k$ is the actual displacement directly evaluated using the atomic displacements between the equilibrium geometry of the $\prescript{1}{}{A}_1$ and the $\prescript{1}{}{E}$ states. Then the HRF can be evaluated as
\begin{equation}
\begin{aligned}
    S_{k,F} &= \frac{\omega_k \Delta Q_{k,F}^2}{2\hbar}\\
    &= \frac{\omega_k}{2\hbar} \left(\dfrac{\widetilde{\mathbf{F}}_k}{\omega_k^2}\right)^2 
    \\
    &= \frac{\omega_k \Delta Q_k^2}{2\hbar} \left( 1 + \frac{3a_{k3}}{2a_{k2}} \Delta Q_k \right)^2 \\
    &= S_k \left(1 + 3\frac{a_{k3}}{a_{k2}}\Delta Q_k + O\left(a_{k3}^2\right) \right) \\
    &\approx S_k (1 + 3r_k),
\end{aligned}
\end{equation}
which is exactly the anharmonic HR factor $S_{k,\text{an}}$ (equation~(\ref{s-eq:an-hrf})). In the last step we neglected the term containing $a_{k3}^2$. Therefore, by evaluating the HRFs using $\Delta Q_{k,F}$ computed using equation~(\ref{eq:QkF}), we automatically include anharmonic effects to the lowest order. We also note that if anharmonic effects are negligible, then by definition $S_{k,F} = S_k$.

Supplementary Table~\ref{s-tab:hrf} reports the total HRFs computed using either forces or displacements. We also report the unitless ratio $r$ of equation~(\ref{eq:r}) evaluated based on the cubic fitting of the PECs of the $\prescript{1}{}{A}_1$ along the one dimensional configurational coordinate diagram connecting the equilibrium geometries of the $\prescript{1}{}{A}_1$ and $\prescript{1}{}{E}$ states. By comparing $\frac{\sum_k S_{k,F}}{\sum_k S_k}$ with $(1+3r)$, we can conclude that anharmonic effects are included when calculating HRFs using $\Delta Q_{k,F}$.

\bibliographystyle{unsrt}

\providecommand{\noopsort}[1]{}\providecommand{\singleletter}[1]{#1}%